\documentclass[%
 reprint,
 amsmath,amssymb,
 aps,
]{revtex4-2}
\usepackage{xcolor}
\usepackage{graphicx}% Include figure files
\usepackage{dcolumn}% Align table columns on decimal point
\usepackage{bm}% bold math
\usepackage{hyperref}

\begin{document}

\preprint{APS/123-QED}

\title{Phonon trapping states as a witness for generation of phonon blockade in a hybrid micromaser system}

\author{Hugo Molinares$^{1}$}
\email{hugo.molinares@mayor.cl}
\author{Vitalie Eremeev$^{2,3}$}
\email{vitalie.eremeev@udp.cl}
\author{Miguel Orszag$^{1,4}$}
\email{corresponding author: miguel.orszag@umayor.cl}

\affiliation{
$^{1}$Centro de Optica e Informaci\'on Cu\'antica, Universidad Mayor,\\
camino la Piramide 5750, Huechuraba, Santiago, Chile
}
\affiliation{
$^{2}$Instituto de Ciencias B\'asicas, Facultad de Ingenier\'ia y Ciencias, Universidad Diego Portales, Av.
 Ejercito 441, Santiago, Chile
}
\affiliation{
$^{3}$Institute of Applied Physics, Academiei 5, MD-2028, Chi\c{s}in\u{a}u, Moldova
}
\affiliation{
$^{4}$Instituto de F\'isica, Pontificia Universidad Cat\'olica de Chile, Casilla 306, Santiago, Chile
}
\date{\today}

\begin{abstract}
In a hybrid micromaser system consisting of an optical cavity with a moving mirror connected to a low temperature thermal bath, we demonstrate, both analytically and numerically, that for certain interaction times between a random atomic flux and the optomechanical cavity, vacuum phonon trapping states %\cite{Orszag2022} 
are generated. Furthermore, under the approach of the master equation with independent phonon and photon thermal baths, we show that the trapping of the phonons and photons is achieved for the same interaction times. The results also indicate that by increasing the cavity-oscillator coupling one may generate a coherent phonon state aside from the trapping states. Within the same hybrid system, but now connected to the squeezed phonon reservoir, a phonon blockade effect can be engineered. Moreover, we identify an interconnection between the trapping and blockade effects, particularly if one approaches the vacuum trapping state, strong phonon blockade can be achieved when the system is connected with a weakly squeezed phonon reservoir.
\end{abstract}

\maketitle

\section{Introduction}\label{sec1}

There are many interesting and important applications of quantum effects realized in cavity QED systems \cite{HR}, which commonly consists in a Fabry-Perot cavity (FPC) interacting with atoms, and the whole system interchanging its energy to a reservoir, usually a thermal one. In the mentioned system the standard interaction between the cavity mode and the atom is described by the well-known Jaynes-Cummings model \cite{JC}, and some external sources can be involved additionally to drive the cavity mode and pump the atoms in order to stimulate some expected effects. For example, maser/laser radiation \cite{Scully, Orszag}, superradiance \cite{Dicke, Gross}, Schrodinger cat states \cite{HR}, photon squeezing \cite{Walls, Scully}, sub-Poissonian photon statistics and “trapping” states \cite{Filipowicz,Weidinger,Walther2001, Meystre}, are some of such quantum phenomena.

An interesting at fundamental level, as well rich in quantum features, is the micromaser (MM) model \cite{Scully,Orszag}, widely investigated theoretically \cite{Filipowicz,2phMM,Bergou} and experimentally \cite{Maser,2phMexp, Walther, Weidinger} in the decades of 80’s and 90’s. Actually, the most attractive quantum properties for applications observed in MM, are related to the photon blockade \cite{Walther2001}, squeezing \cite{Dag2016}, non-classical states \cite{Nation2013,Dambach2019,Kouzelis2020}, MM synchronization \cite{Davis2016}. Having in mind these MM’s features, we were motivated to investigate a kind of hybrid micromaser (HMM) architecture, particularly considering an oscillating mirror of the FPC, so including additionally the optomechanical interaction in the system’s Hamiltonian \cite{Aspelmeyer}. As result of this, one expects that the HMM will enrich the quantum features as compared to the standard MM because of the influence of a new degree of freedom, that is the mechanical motion mode. The hybrid systems considering interactions between cavities, atoms and mechanical resonators, nowadays are of great importance for many applications in quantum technologies \cite{Aspelmeyer, Mun2018,Mir2020} as well are the promising candidates to probe the foundations of the quantum mechanics, e.g. macroscopic quantum superpositions \cite{Nak1999, Liao2016, Mon2017, Teh2018}, quantum correlations at macroscopic scales \cite{Sca2013, Mar2018}, phonon and photon  blockade effects \cite{Liu2010, Didier2011, Wang2016, Restrepo2017, Xu2019, expPB, Shi2019, Zheng2019, Yang2020, Lin2021, APL2021}.

In the present work we develop a comprehensive investigation of the effects of trapping and blockade for phonons and photons in the HMM model. On the one hand, from the fundamental point of view, it is important to know how the mechanical degree of freedom affects the photon quantum features particular to the standard MM. On the other hand, for practical reasons, it is of high interest to engineer tools of control of different elements in a system, and the hybrid one is a marvel candidate for quantum control protocols. Therefore, we embrace these two strategies that converge to results of fundamental and application importance.

Hence, in the HMM system, connected to a low temperature thermal reservoir, we show the possibility of generation of the vacuum phonon trapping states. Besides, the trapping of the phonons and photons can be achieved for the same interaction times. Another quantum effect which attracted our interest is the phonon blockade (antibunching). The effects of phonon and photon blockade were intensively studied the last decade, particularly in the hybrid setups \cite{Liu2010, Didier2011, Wang2016, Restrepo2017, Xu2019, expPB, Shi2019, Zheng2019, Yang2020, Lin2021, APL2021}. In order to realize such an effect commonly one needs a quantum strong nonlineariry in the system, i.e. for the photon blockade it can be a strong Jaynes-Cummings interaction \cite{Imamoglu97}, strong optomechanical coupling \cite{Rabl2011}, Kerr-type nonlinearity \cite{Didier2011}; for the phonon blockade could be a strong effective atom-cavity-mechanics interaction \cite{Restrepo2017}, strong opto/spin-mechanical coupling \cite{Liu2010, Xu2019} or the nonlinear phonon term similar to the Kerr one \cite{APL2021}. An alternative mechanism to stimulate the blockade effect, which is known as unconventional, is based on the destructive quantum interference between different paths so that the two-quanta and higher order excitations are blocked, e.g. photon \cite{Liew2010, Majumdar2012} and phonon \cite{Yang2020, Lin2021} unconventional blockade. 
Our proposal of HMM is developed in a weak optomechanical coupling regime and hence to enhance the quantum nonlinearity in the HMM we propose to connect the mechanical oscillator (MO) to a squeezed phonon reservoir. Therefore, by connecting the HMM to the squeezed phonon reservoir, one may stimulate and control the phonon blockade effect. Particularly, in the case of weak squeezed phonon reservoir and weak optomechanical coupling one observes the creation of the phonon blockade when the system approaches the vacuum phonon trapping state. As result, we can control one quantum effect by the other one as tuning some system’s parameters. Such protocol could be very useful for phonon blockade detection, which from the experimental point of view is challenging as pointed in the theoretical and experimental proposals \cite{Didier2011,Wang2016,expPB,APL2021}. Therefore, in this work we suggest to detect the phonon blockade effect by using the prototype of the photon trapping experimental setup \cite{Weidinger} adjusted (designed) for the HMM.

This work is organized as follows. In Sec. \ref{sec:level2} we present the conceptual model of the HMM, by defining the Hamiltonian and the unitary dynamics. In Sec. \ref{sec:level3} the master equation for HMM is developed to get the phonon and photon density operators under the decoherence effects. Next, in Secs. \ref{sec:level4} and \ref{sec:level5} we demonstrate how the trapping states of the phonon and photons are realized, and the effect of synchronization between these trapping states is shown. Section \ref{sec:level6} is devoted to the study of the second order coherence function by which the maser effect is analyzed as function of the optomechanical coupling. In Sec. \ref{sec:level7} we present the effect of phonon blockade and explain how it is controlled by the vacuum squeezed reservoir and the optomechanical coupling. Finally, Sec. \ref{sec:level8} is devoted to Discussion, where we analyze the obtained results, suggest an experimental setup and make the conclusions.

%%%%%%%%%%%Figure 1%%%%%%%%%%%%%%
\begin{figure}[t]
\centering
\includegraphics[width=1\linewidth]{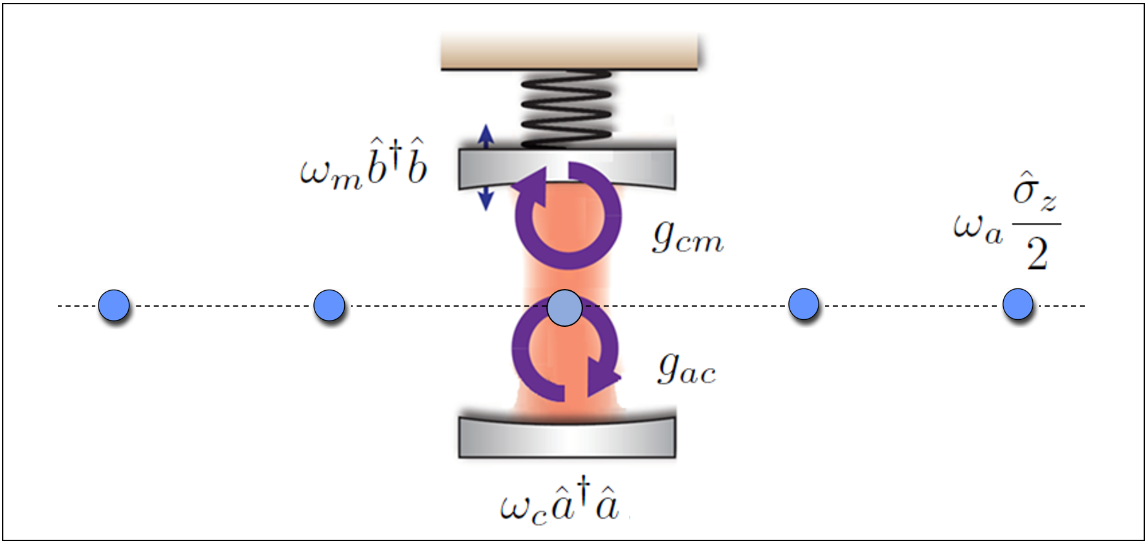}
\caption{Schematic diagram of a hybrid micromaser model, where a set of atoms (one at a time) crosses the cavity field connected to a mechanical oscillator.}
\label{fig1}
\end{figure}
%%%%%%%%%%%%%%%%%%%%%%%%%%%%%%%%%%

%%%%%%%%%%%%%%%%%%%%%%%%%%%%%%%%%%
%%%%%%%%%%%%%%%%%%%%%%%%%%%%%%%%%%%%%%%%%% %%%%%%%%%%%%%%%%%%%%%%%%%%%%%%%%%%%%%%%%%%%%%%%%%%%
\section{MODEL AND MASTER EQUATION OF THE HYBRID MICROMASER}

\subsection{\label{sec:level2}Hamiltonian of the atom-cavity-mechanics system}
%%%%%%%%%%%%%%%%%%%%%%%%%%%%%%%%%%%%%%%%%%%%%%%%%%%
Let us consider a hybrid atom-cavity-mechanics system, as illustrated in Fig. \ref{fig1}. The Jaynes-Cummings interaction between the two-level atom of frequency $\omega_{a}$ and the mode of the cavity field of frequency $\omega_{c}$ is quantified by the coupling constant $g_{ac}$. The optomechanical interaction between the cavity and MO of frequency $\omega_{m}$ is quantified by the coupling constant $g_{cm}$. Therefore, we will consider our hybrid system described by the Hamiltonian under the rotating wave approximation as (with $\hbar=1$)
\begin{eqnarray}\label{base}
    \mathcal{\hat{H}}&=&\omega_{a}\frac{\hat{\sigma}_{z}}{2}+\omega_{c}\hat{a}^{\dagger}\hat{a}+\omega_{m}\hat{b}^{\dagger}\hat{b}\nonumber\\
    &+& g_{ac}\left(\hat{a}\hat{\sigma}_{+}+\hat{a}^{\dagger}\hat{\sigma}_{-}\right)
    -g_{cm}\hat{a}^{\dagger}\hat{a}\left(\hat{b}^{\dagger}+\hat{b}\right),
\end{eqnarray}
where $\hat{a}(\hat{a}^{\dagger})$ and $\hat{b}(\hat{b}^{\dagger})$ are the annihilation (creation) operators of the cavity and the MO modes, respectively. These operators obey the boson commutation relation $[\hat{a}(\hat{b}), \hat{a}^{\dagger}(\hat{b}^{\dagger})]=1$; and $\hat{\sigma}_{z}, \hat{\sigma}_{+}$, $\hat{\sigma}_{-}$ are respectively, the $z$-Pauli, raising and lowering spin operators.

In the following, we calculate the Hamiltonian in the interaction picture (rotating at the mechanical frequency $\omega_{m}$)  
\begin{equation}\label{model}
    \hat{\tilde{\mathcal{H}}}=\delta\frac{\hat{\sigma}_{z}}{2}+g_{ac}\left(\hat{\sigma}_{+}\hat{a}e^{i \mathcal{\hat{F}}(t)}+\hat{\sigma}_{-}\hat{a}^{\dagger}e^{-i \mathcal{\hat{F}}(t)}\right),
\end{equation}
where we used the Hermitian operator $\mathcal{\hat{F}}(t)=-i\lambda\left(\hat{b}^{\dagger}\eta-\hat{b}\eta^{*}\right)$, with $\lambda=g_{cm}/\omega_{m}$, $\eta=e^{i\omega_{m}t}-1$ and $\delta=\omega_{a}-\omega_{c}$ is the detuning (see details in Appendix \ref{apendice1}).

To solve the quantum dynamics, we proceed to derive the time evolution operator for the Hamiltonian in Eq. \ref{model} defined by  $\mathcal{\hat{U}}(t)=\sum_{n=0}^{\infty}(-i t \mathcal{\hat{\tilde{H}}})^{n}/n!$. By a straightforward calculation, we get the time evolution operator: 
\begin{equation}\label{unitary}
\mathcal{\hat{U}}(t)=
\begin{pmatrix}
\mathcal{\hat{C}} & -ig_{ac}\mathcal{\mathcal{\hat{S}}}\hat{a}e^{i\mathcal{\hat{F}}}\\\\
 -ig_{ac}e^{-i\mathcal{\hat{F}}}\hat{a}^{\dagger}\mathcal{\mathcal{\hat{S}}} & \mathcal{\hat{D}}
\end{pmatrix},
\end{equation}
where
\begin{align}
    \mathcal{\hat{C}}&=\cos{\left(t\sqrt{\hat{\varphi}+g_{ac}^{2}}\right)}-\frac{i\delta}{2}\frac{\sin{\left(t\sqrt{\hat{\varphi}+g_{ac}^{2}}\right)}}{\sqrt{\hat{\varphi}+g_{ac}^{2}}},\\
    \mathcal{\hat{D}}&=\cos{\left(t\sqrt{\hat{\varphi}}\right)}
    +\frac{i\delta}{2}\frac{\sin{\left(t\sqrt{\hat{\varphi}}\right)}}{\sqrt{\hat{\varphi}}}\label{eq5},\\
    \mathcal{\mathcal{\hat{S}}}&=\frac{\sin{\left(t\sqrt{\hat{\varphi}+g_{ac}^{2}}\right)}}{\sqrt{\hat{\varphi}+g_{ac}^{2}}},
\end{align}
where $\hat{\varphi}=g_{ac}^{2}\hat{a}^{\dagger}\hat{a}+\left(\delta/2\right)^{2}$. According to the last experimental results, the optomechanical coupling cover a wide spectrum of values \cite{Aspelmeyer}. However, in the Eq. \ref{unitary} we have neglected the fast oscillations of the mechanical frequency, within which $\eta=-1$ is valid in the weak coupling regime, $g_{cm}\ll\omega_{m}$ \cite{Murch,Painter,Xuereb}.

%%%%%%%%%%%%%%%%%%%%%%%%%%%%%%%%%%%%%%%%%%%%%%%%%%%
\subsection{\label{sec:level3}Master equation of HMM}
%%%%%%%%%%%%%%%%%%%%%%%%%%%%%%%%%%%%%%%%%%%%%%%%%%%
In the following we study how the optomechanical interaction affects the well known micromaser model \cite{Scully,Orszag}. Let us consider the initial atom-cavity-MO operator $\hat{\rho}(0)=\hat{\rho}_{a}(0)\otimes\hat{\rho}_{c}(0)\otimes\hat{\rho}_{m}(0)$, then after the interaction time $\tau$, the density operator becomes
\begin{equation}
    \hat{\rho}(\tau)=\mathcal{\hat{U}}(\tau)
\left[\hat{\rho}_{a}(0)\otimes \hat{\rho}_{c}(0)\otimes\hat{\rho}_{m}(0)\right]\mathcal{\hat{U}}^{\dagger}(\tau),
\end{equation}
where the evolution operator $\mathcal{\hat{U}}(\tau)$ is defined in Eq. \ref{unitary}. Now, in order to study the dynamics of the mechanical subsystem, we trace over the atom and cavity field subsystems. For this, we assume that initially the atom is in ground state, i.e. $\hat{\rho}_{a}(0)=\vert0\rangle\langle0\vert$, and the cavity field is in a coherent state
\begin{equation}\label{coh}
    \hat{\rho}_{c}(0)=e^{-\vert\alpha\vert^{2}}\sum_{n,m}\frac{\alpha^{n}\alpha^{*m}}{\sqrt{n!}\sqrt{m!}}\vert n\rangle\langle m\vert ,
\end{equation}
where $\alpha$ is, in general, a complex number. By performing the matrix product and tracing over the atom, we get
\begin{equation}\label{rocm}
    \hat{\rho}_{c,m}(\tau)
    =
    \hat{\rho}_{m}(0)\mathcal{\hat{D}}\hat{\rho}_{c}(0)\mathcal{\hat{D}}^{\dagger}+
    g^{2}_{ac}e^{i\mathcal{\hat{F}}}\hat{\rho}_{m}(0)e^{-i\mathcal{\hat{F}}}\mathcal{\mathcal{\hat{S}}}\hat{a}\hat{\rho}_{c}(0)\hat{a}^{\dagger}\mathcal{\mathcal{\hat{S}}}.
\end{equation}
Finally, tracing over the cavity density operator, so the atom and cavity fields leave the MO in the state defined by the reduced density operator
\begin{eqnarray}\label{rom}
    \hat{\rho}_{m}(\tau)&=&A(\tau)\hat{\rho}_{m}(0) + B(\tau)e^{i \mathcal{\hat{F}}}\hat{\rho}_{m}(0)e^{-i \mathcal{\hat{F}}}\nonumber\\
    &\equiv&
    \hat{\mathcal{M}}_{m}(\tau)\hat{\rho}_{m}(0),
\end{eqnarray}
where $\hat{\mathcal{M}}_{m}$ is the gain superoperator acting on $\hat{\rho}_{m}$, and the coefficients $A(\tau)$ and $B(\tau)$ are given by
\begin{eqnarray}\label{coefab}
    A(\tau)
    &=&
    e^{-\vert \alpha\vert ^{2}}\sum_{n}\frac{\vert \alpha\vert ^{2n}}{n!}\nonumber\\
    &\times&
    \left[\cos^{2}{\left(\tau\sqrt{\varphi_{n}}\right)}+\frac{\delta^{2}}{4}\frac{\sin^{2}{\left(\tau\sqrt{\varphi_{n}}\right)}}{\varphi_{n}}\right],\\
    B(\tau)
    &=&
    e^{-\vert \alpha\vert ^{2}}\sum_{n}\frac{\vert \alpha\vert ^{2(n+1)}}{n!}\frac{g_{ac}^{2} \sin^{2}{\left(\tau\sqrt{\varphi_{n+1}}\right)}}{\varphi_{n+1} }\label{coefab2},
\end{eqnarray}
with $\varphi_{n+1}=g_{ac}^{2}(n+1)+\left(\delta/2\right)^{2}$. Since $\emph{Tr}\left\{\hat{\rho}_{m}(\tau)\right\}=1$ we can see that it is fulfilled $A(\tau)+B(\tau)=1$. The Eqs. \ref{rom}, \ref{coefab}  and  \ref{coefab2}
are the main results of this section (see details in Appendix \ref{apendice2}).

We assume that initially there is no interaction between the atom and the MO, see Eq. \ref{base}. However, the tripartite system produces an atom-MO interaction mediated by the cavity \cite{Orszag2022}.
In the following, considering the generation of this indirect qubit-MO interaction, we develop the ME for the generalized pump statistics \cite{Orszag, Scully}. We first assume that the $j-th$ atom is “injected” at the time $t_{j}$ into the cavity. Then, the MO density operator after the cavity field interacted with the $j-th$ atom, can be written as $\hat{\rho}_{m}(t_{j}+\tau)=\hat{\mathcal{M}}_{m}(\tau)\hat{\rho}_{m}(t_{j})$. Now, if $k$ atoms are excited, then $\hat{\rho}_{m}(t)=\hat{\mathcal{M}}^{k}_{m}(\tau)\hat{\rho}(0)$.

Of course, while the number of excited atoms is not known, one may use the probability defined as $P(k)=C_{Kk}p^{k}(1-p)^{K-k}$, where $C_{Kk}=K!/k!(K-k)!$, $p$ is the probability each atom has of being excited and $K$ is the total number of atoms involved in the lasing process, i.e., $0<k<K$. Therefore, the average number of atoms contributing to the gain is $\langle k\rangle=pK$. 

%%%%%%%%%%%Figure 2%%%%%%%%%%%%%%
\begin{figure*}[t]
\centering
\includegraphics[width=0.245\linewidth]{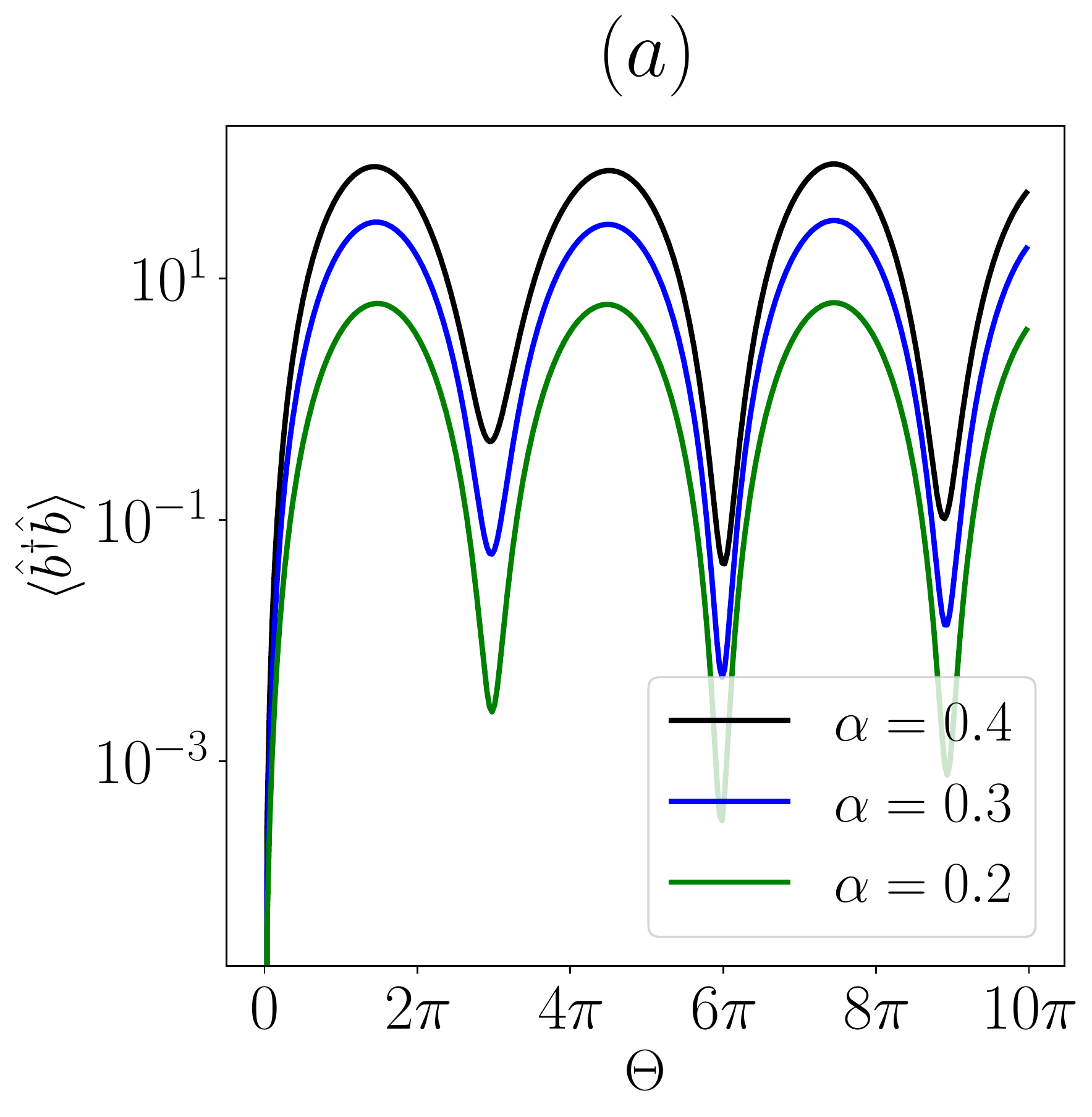}
\includegraphics[width=0.245\linewidth]{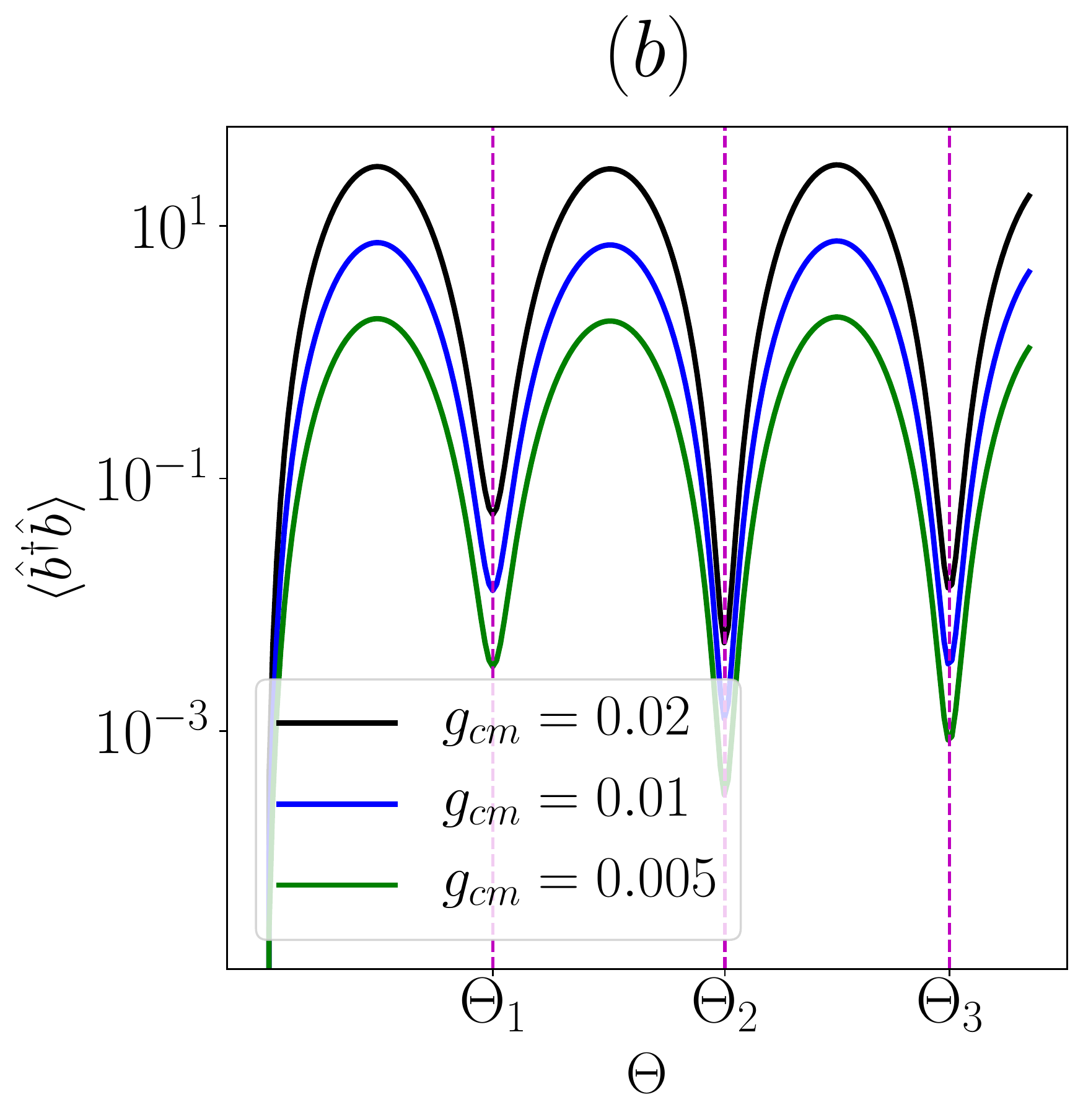}
\includegraphics[width=0.245\linewidth]{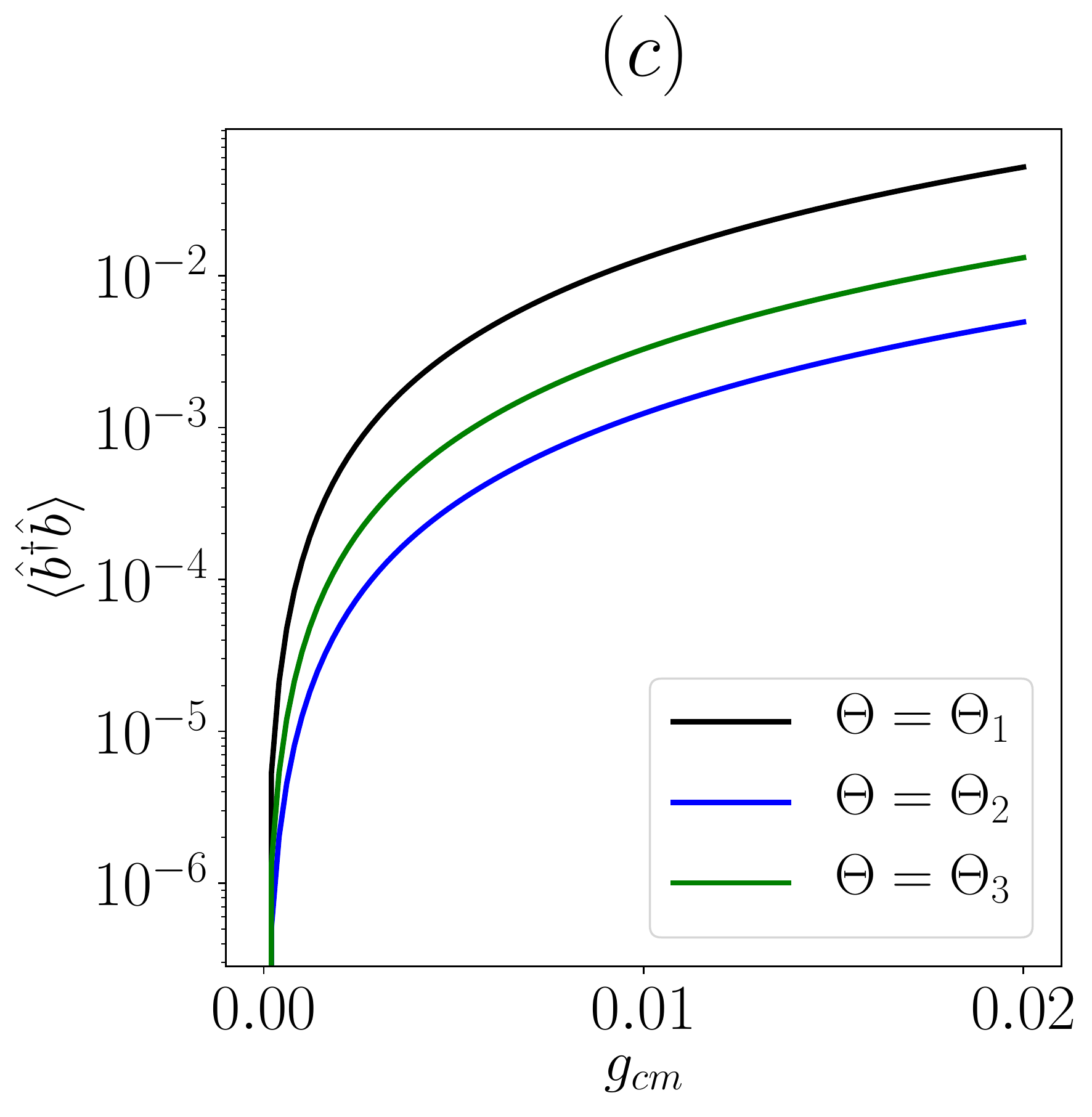}
\includegraphics[width=0.245\linewidth]{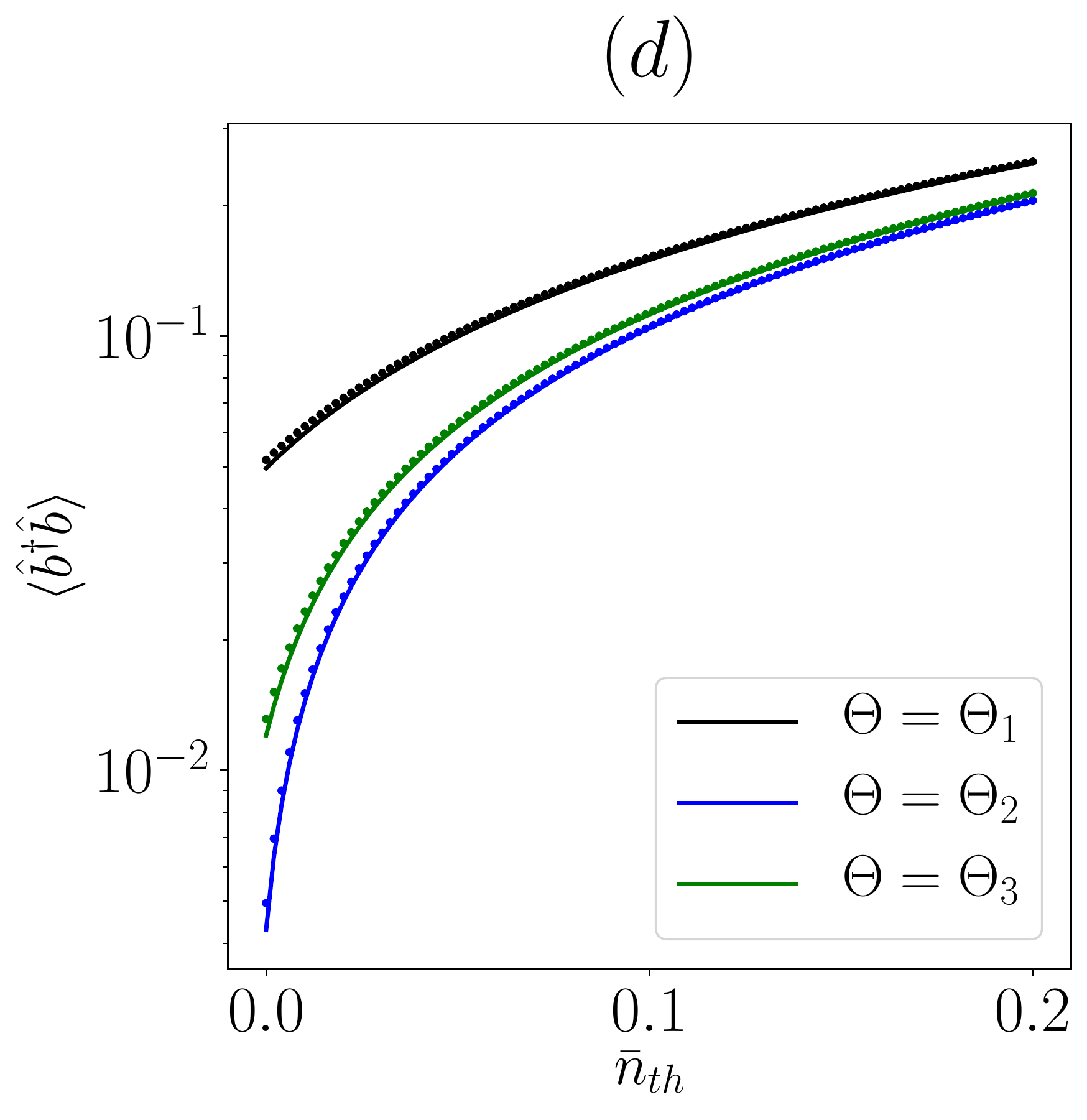}
\caption{Steady state average phonon number as a function of the pump parameter $\Theta$ for different: $(a)$ cavity initial coherent states, $\alpha$ (with $g_{cm}=0.02$, $\bar{n}_{th}=0$) and $(b)$ optomechanical couplings, $g_{cm}$ (with $\alpha=0.3$, $\bar{n}_{th}=0$). Steady state average phonon number calculated in the minimum value for $\Theta = \left\{\Theta_{1}\approx9.32, \Theta_{2}\approx18.81, \Theta_{3}\approx28.01\right\}$ and by varying: $(c)$ optomechanical coupling, $g_{cm}$ (with $\alpha=0.3$, $\bar{n}_{th}=0$), and $(d)$ thermal bath temperature, $\bar{n}_{th}$ (with $g_{cm}=0.02$, $\alpha=0.3$): $(i)$ Using analytical expression (\ref{analitico}) (solid line) and $(ii)$ using numerical simulation of the ME (\ref{me}) (dotted). Other parameters (in units of $\omega_m$) are: $r=80$, $g_{ac}=3$, $\delta=0$ and $\kappa_{b}=0.05$.}
\label{Fig2}
\end{figure*}
%%%%%%%%%%%%%%%%%%%%%%%%%%%%%%%

In the rest of our work we assume a random arrival of the atoms, that corresponds to $p\rightarrow 0$ in the micromaser notation \cite{Orszag}. Taking into account the above, and considering the gain part assisted by atoms injected at the rate $r$, also including the loss term for the MO subsystem, so the ME reads
\begin{eqnarray}\label{me}
    \frac{d\hat{\rho}_{m}}{dt}
    &=&
    r\left(\hat{\mathcal{M}}_{m}(\tau)-1\right)\hat{\rho}_{m}(t)\nonumber\\
    &+&
    \frac{\kappa_{b}}{2}(1+\bar{n}_{th})\mathcal{L}[\hat{b}]\hat{\rho}_{m}(t)+\frac{\kappa_{b}}{2}\bar{n}_{th}\mathcal{L}[\hat{b}^{\dagger}]\hat{\rho}_{m}(t),
\end{eqnarray}
where $r$ is the rate at which atoms enters into the
cavity and $\mathcal{L}[\hat{b}]=2\hat{b}\hat{\rho} \hat{b}^{\dagger}-\hat{b}^{\dagger}\hat{b}\hat{\rho}-\hat{\rho} \hat{b}^{\dagger}\hat{b}$ is the standard Lindbladian describing the decoherence effect. Here $\kappa_{b}$ is the decay rate of the mechanical mode to the phonon thermal bath with $\bar{n}_{th}\equiv(\exp{[\hbar\omega_m/k_B T_m]}-1)^{-1}$ at temperature $T_m$. 

%%%%%%%%%%%%%%%%%%%%%%%%%%%%%%%%%%%%%%%%%%%%%%%%%%%
\section{\label{sec:level4}Generation of trapping states in the Hybrid micromaser}
%%%%%%%%%%%%%%%%%%%%%%%%%%%%%%%%%%%%%%%%%%%%%%%%%%%
\subsection{\label{sec:level4}Phonon trapping states}

In terms of Glauber’s $P$-distribution, the MO density matrix can be written as $\hat{\rho}_{m}(t)=\int d^{2}\beta P(\beta,\beta^{*},t)\vert \beta\rangle\langle\beta\vert $. The gain superoperator $\mathcal{\hat{M}}_{m}(\tau)$ in this notation generates
\begin{eqnarray}\label{gen}
    \mathcal{\hat{M}}_{m}(\tau)\hat{\rho}_{m}(t)&=&\int d^{2}\beta P(\beta,\beta^{*},t)\{A(\tau)|\beta\rangle\langle\beta|
    \nonumber\\
    &+&B(\tau)|\beta+\lambda\rangle\langle\beta+\lambda|\}.
\end{eqnarray}
As we see, the first term of the gain is proportional to the initial coherent state, and in the second term, the coherent state is modified due to the displacement operator. 

A witness for the vacuum phonon trapping states correspond to a sharp decrease in the steady state of the average phonon number, that can be calculated solving the ME (Eq. \ref{me}). 

By the standard technique to convert the ME into a Fokker-Planck second-order differential equation \cite{Carmichael} with the loss term $\mathcal{L}\hat{\rho}_{m}=\frac{\kappa_{b}}{2}\left(\frac{\partial}{\partial\beta}\beta+\frac{\partial}{\partial\beta^{*}}\beta^{*}\right)P+\kappa_{b}\bar{n}_{th}\frac{\partial^{2} P}{\partial \beta\partial \beta^{*}}$, we get the time-dependent Fokker-Planck equation
\begin{eqnarray}\label{fp}
    \frac{\partial P}{\partial t}
    &=&
    \frac{\partial P}{2\partial \beta}\left(\kappa_{b}\beta-2B(\tau)\lambda r\right)
    +\frac{\partial P}{2\partial \beta^{*}}\left(\kappa_{b}\beta^{*}-2B(\tau)\lambda r\right)\nonumber\\
    &+&\kappa_{b}\bar{n}_{th}\frac{\partial^{2} P}{\partial \beta\partial \beta^{*}}+\left(\kappa_{b}+r\left[A(\tau)+B(\tau)-1\right]\right)P.
\end{eqnarray}
For an initial thermal distribution we find a solution of the Fokker-Planck equation (see details in Appendix \ref{apendice3}):
\begin{equation}\label{sfp}
    P(\beta,\beta^{*},t)=\frac{1}{\pi\bar{n}_{th}}\exp{\left[-\vert \beta-\beta_{1}\vert ^{2}/\bar{n}_{th}\right]},
\end{equation}
with $\beta_{1}=\frac{2\lambda r B(\tau)}{\kappa_{b}}(1-\exp{\left[-\kappa_{b}t/2\right]})$. Now, using the definition for the average phonon number, $\langle \hat{b}^{\dagger}\hat{b}\rangle\equiv Tr\left\{\hat{b}^{\dagger}\hat{b}\hat{\rho}_{m}(t   )\right\}$ we get
\begin{equation}
    \langle \hat{b}^{\dagger}\hat{b}\rangle
    =
    \bar{n}_{th}+\frac{4B^{2}(\tau)\lambda^{2}r^{2}}{\kappa^{2}_{b}}\left(1-\exp{\left[-\kappa_{b}t/2\right]}\right)^{2}.
\end{equation}
Hence, the steady state $(t\rightarrow\infty)$ average phonon number is
\begin{equation}\label{analitico}
    \langle \hat{b}^{\dagger}\hat{b}\rangle
    =\bar{n}_{th}+\frac{4B^{2}(\tau)\lambda^{2}r^{2}}{\kappa^{2}_{b}}.
\end{equation}

Although the above expression can be written in terms of an apparently simple formula, the state of the system changes dramatically as a function of the atomic flow rate $r$ and atom-cavity interaction time, parametrized by the pump parameter $\Theta=\tau \left(\omega_{m}r\right)^{1/2}$. 

In panels $(a)$ and $(b)$ of Fig. \ref{Fig2} we show the evolution of the average phonon occupation number as a function of $\Theta$ for different amplitudes of the coherent state of the cavity, $\alpha$, and for different optomechanical couplings, $g_{cm}$, respectively. Analytically, the trapping states can be obtained from the condition (see details in Appendix \ref{apendice4})

\begin{equation}\label{cond}
    \sum_{n}\frac{\vert \alpha\vert ^{2(n+1)}\sqrt{n+1}}{(n+1)!}\sin{\left(2g_{ac}\Theta\sqrt{\frac{n+1}{\omega_{m}r}}\right)}=0.
\end{equation}
The above equation can be solved numerically, and therefore, the values of $\Theta$ corresponding to the minimum values of $\langle\hat{b}^{\dagger}\hat{b}\rangle$ are determined, see vertical lines in  Fig. \ref{Fig2}(b).

In Fig. \ref{Fig2}(c) we plot the average phonon number in its minimum, for $\Theta = \left\{\Theta_{1}, \Theta_{2}, \Theta_{3}\right\}$ as function of the optomechanical coupling, $g_{cm}$. Therefore, the optomechanical coupling, although it does not control the critical point, $\Theta$ of the trapping states, it does contribute directly in the mean number of phonons, as results in Eq. \ref{analitico} through the parameter $\lambda$.

In Fig. \ref{Fig2}(d) we show the average phonon occupation number as a function of the temperature of the thermal bath, $\bar{n}_{th}$. We observe that the vacuum trapping phonon states tend to occur for low value of the mean number of thermal phonons. Since $B(\tau)\propto\sqrt{\mid\langle \hat{b}^{\dagger}\hat{b}\rangle-\bar{n}_{th}\mid}$, %$B(\tau)\in\mathbb{R}$. 
so, in order to have a vacuum trapping state, i.e. $\langle\hat{b}^{\dagger}\hat{b}\rangle\rightarrow0$, the thermal phonon number must be close to zero \cite{Filipowicz}.
%%%%%%%%%%%Figure 3%%%%%%%%%%%%%%
\begin{figure*}[t]
\centering
\includegraphics[width=0.35\linewidth]{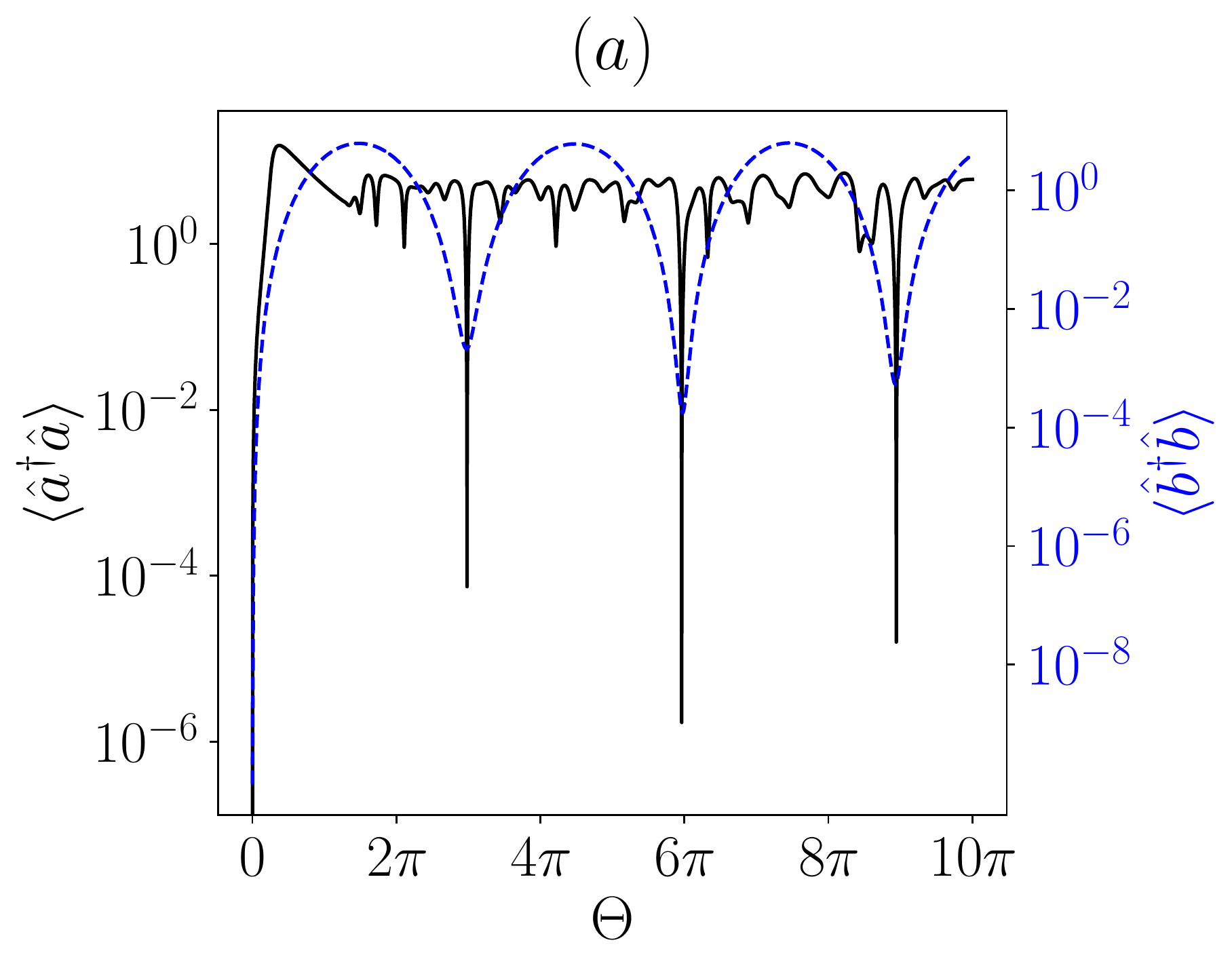}
\includegraphics[width=0.27\linewidth]{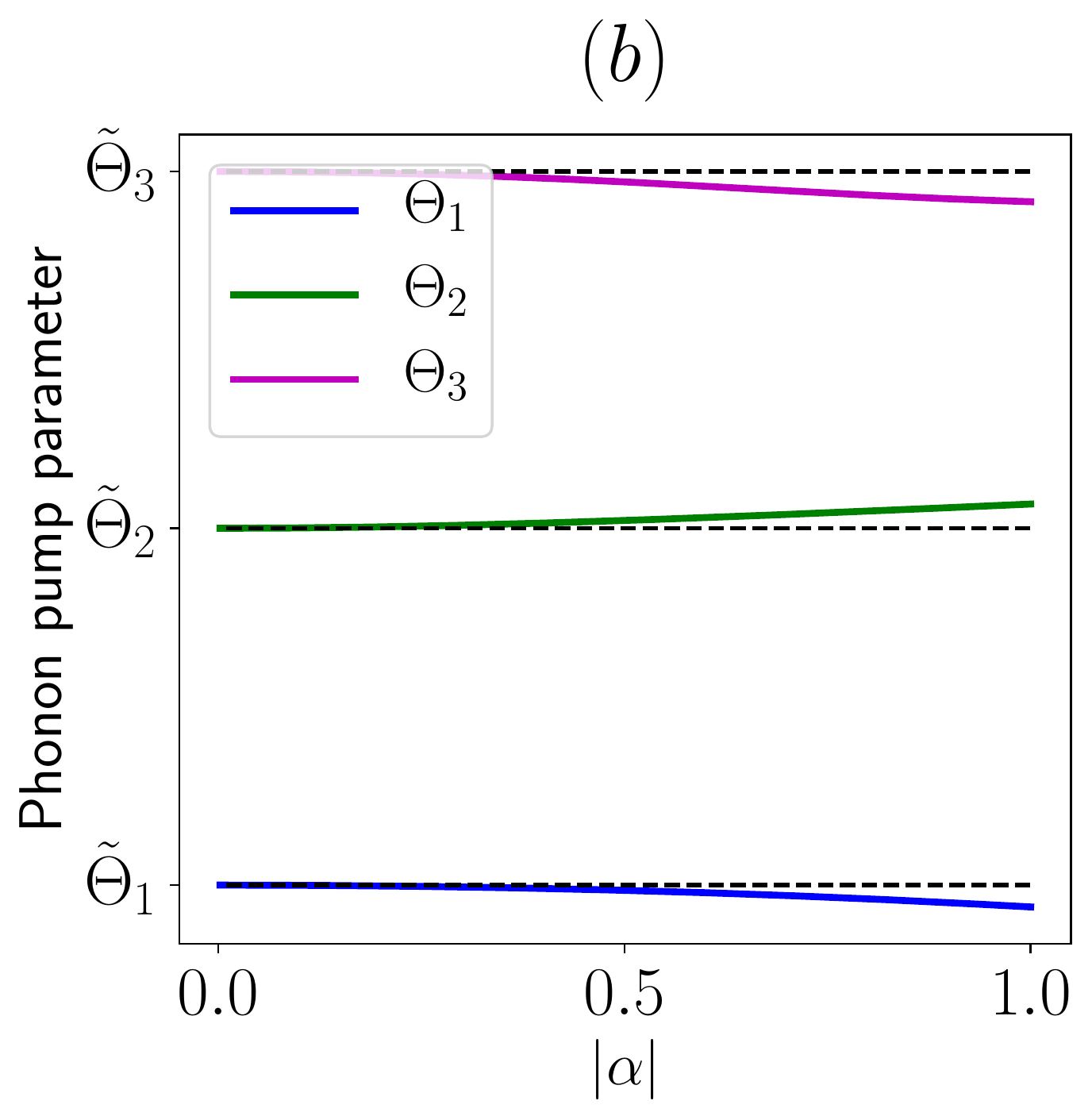}
\includegraphics[width=0.367\linewidth]{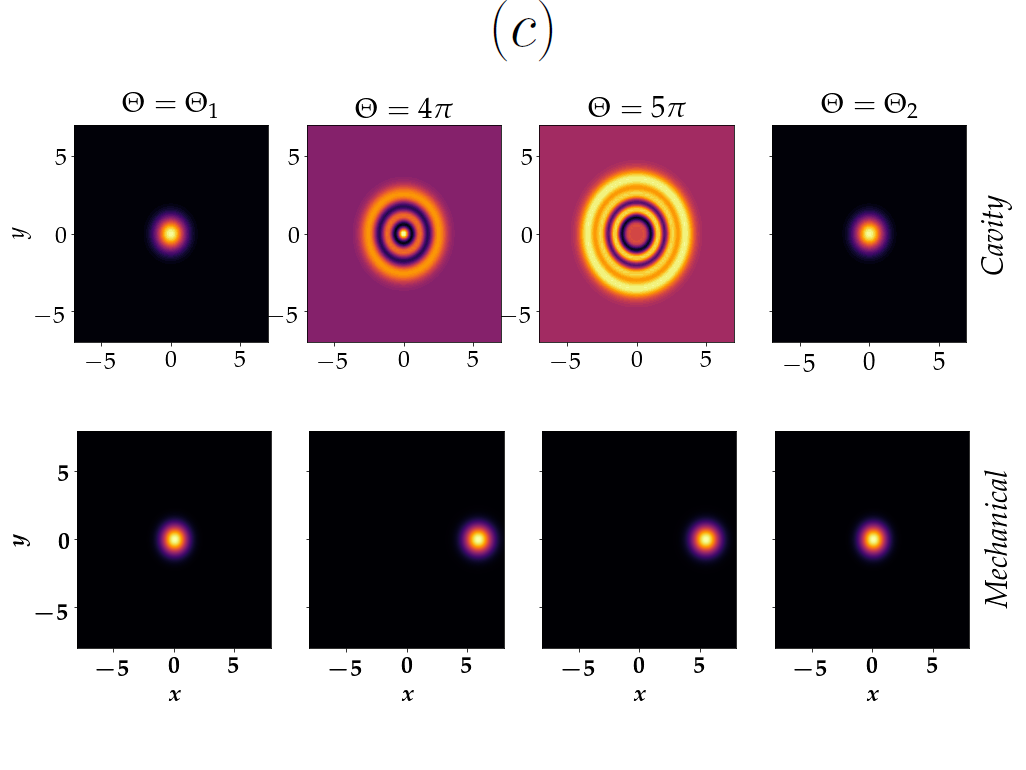}
\caption{$(a)$ Steady state average occupation number of photons [black-solid line, using Eq. \ref{me2}] and phonons [blue-dashed line, using Eq. \ref{analitico}] as a function of the pump parameter $\Theta$. $(b)$ Phonon pump parameters, $\left\{\Theta_{1}, \Theta_{2}, \Theta_{3}\right\}$ (corresponding to vacuum phonon trapping states) as a function of the amplitude of coherent state, $\vert \alpha\vert $ (using Eq. \ref{cond}). The horizontal dashed lines $\left\{\tilde{\Theta}_{1}, \tilde{\Theta}_{2}, \tilde{\Theta}_{3}\right\}$ correspond to the values for the vacuum photon trapping states. As observed, the vacuum photon and phonon trapping states are getting closer for $\vert \alpha\vert \ll1$. $(c)$ Visualization of Wigner function for the cavity and MO density operators at different interaction times. As expected, for $\left\{\Theta_{1},\Theta_{2}\right\}$, both modes end up in the vacuum trapping states. The parameters (in units of $\omega_m$) are: $r=80$, $g_{ac}=3$, $g_{cm}=0.02$, $\alpha=0.3$, $\kappa_{a}=3$, $\kappa_{b}=0.05$, $\bar{n}_{th}=0$.}
\label{Fig3}
\end{figure*}
%%%%%%%%%%%%%%%%%%%%%%%%%%%%%%%
%%%%%%%%%%%%%%%%%%%%%%%%%%%%%%%%%%%%%%%%%%%%%%%%%%%
\subsection{\label{sec:level5}Photon trapping states and synchronization with the phonon mode}
%%%%%%%%%%%%%%%%%%%%%%%%%%%%%%%%%%%%%%%%%%%%%%%%%%%
In the previous subsection, we studied how the phonon trapping states can be realized by developing the MM model \cite{Orszag} for the HMM with optomechanical coupling, so that the indirect interaction between the MO and atoms is mediated by the cavity. On the other hand, the cavity-atom interaction is direct in this model, via Jaynes-Cummings coupling (see Eq. \ref{base}), then one expects the occurrence of the photon trapping states as in the standard MM. To evaluate how the cavity mode dynamics is affected in the HMM and its correspondence to the mechanics mode, we develop the ME for the cavity field. Following the results in previous section, after tracing over the atom, was obtained the Eq. \ref{rocm}. Next, tracing over the MO and assuming a random arrival of the atoms interacting with the cavity ($p\rightarrow0$), we get the ME for the cavity field:
\begin{eqnarray}\label{me2}
    \frac{d\hat{\rho}_{c}}{dt}
    &=&
    r\left(\hat{\mathcal{M}}_{c}(\tau)-1\right)\hat{\rho}_{c}(t)\nonumber\\
    &+&
    \frac{\kappa_{a}}{2}(1+\bar{n}_{th})\mathcal{L}[\hat{a}]\hat{\rho}_{c}(t)+\frac{\kappa_{a}}{2}\bar{n}_{th}\mathcal{L}[\hat{a}^{\dagger}]\hat{\rho}_{c}(t),
\end{eqnarray}
where $\kappa_{a}$ is the decay rate of the cavity mode to the bath with $\bar{n}_{th}$ photons on average. In Eq. \ref{me2} $\mathcal{\hat{M}}_{c}(\tau)$ is the gain superoperator acting on $\hat{\rho}_{c}$, defined by
\begin{eqnarray}
    \hat{\mathcal{M}}_{c}(\tau)
    &=&
    g_{ac}^{2}\hat{a}^{\dagger}\frac{\sin{\left(\tilde{\Theta}\sqrt{\hat{\varphi}/\omega_{m}r}\right)}}{\sqrt{\hat{\varphi}}}\hat{\rho}_{c}(0)
    \frac{\sin{\left(\tilde{\Theta}\sqrt{\hat{\varphi}/\omega_{m}r}\right)}}{\sqrt{\hat{\varphi}}}\hat{a}\nonumber\\
    &+&
    \cos{\left(\tilde{\Theta}\sqrt{\hat{\varphi}/\omega_{m}r}\right)}\hat{\rho}_{c}(0)\cos{\left(\tilde{\Theta}\sqrt{\hat{\varphi}/\omega_{m}r}\right)},
\end{eqnarray}
where $\tilde{\Theta}\equiv \Theta$. The above equation shows that in the weak regime $g_{cm}\ll\omega_{m}$, the initial condition for the MO does not directly affect the dynamics of the cavity field, however, as will be discussed later, the initial condition for the cavity field (Eq. \ref{coh}) modifies the value of the pump parameter $\Theta$, where the phonon trapping states are generated. 
%%%%%%%%%%%Figure 4%%%%%%%%%%%%%%
\begin{figure*}[t] 
\centering
\includegraphics[width=0.49\linewidth]{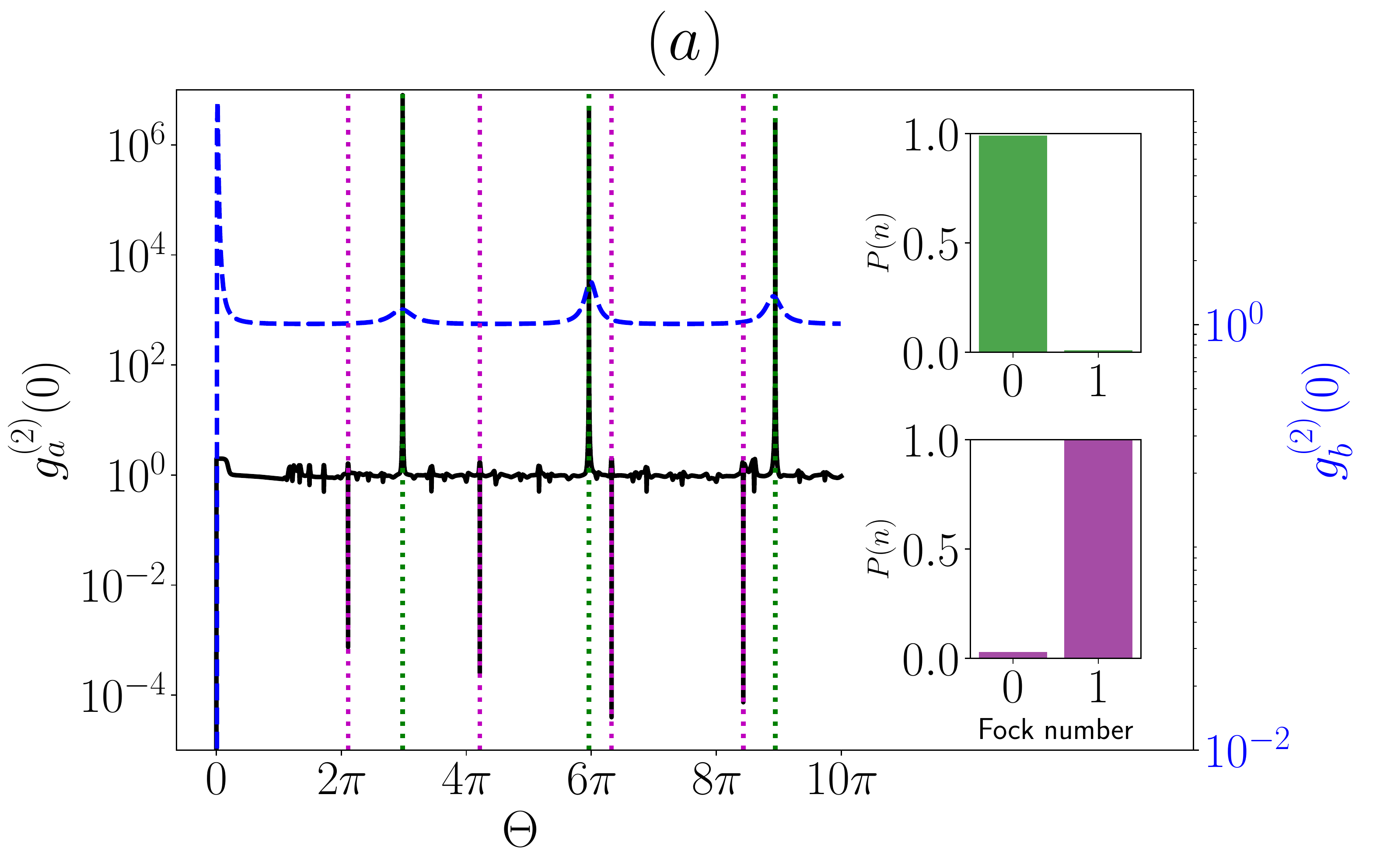}
\hspace{0.3cm}
\includegraphics[width=0.303\linewidth]{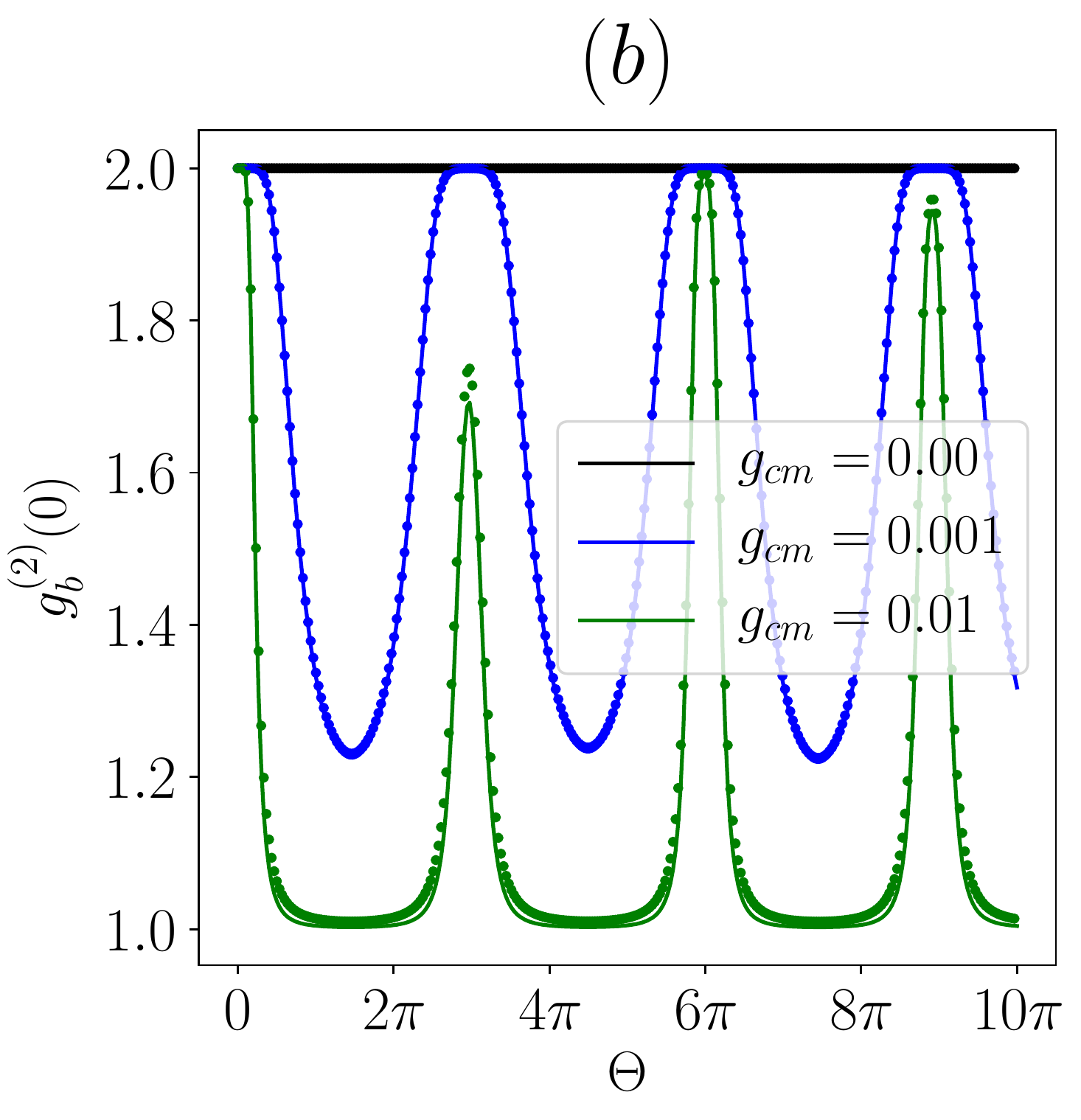}
\caption{$(a)$ Second-order coherence degree for photons (black-solid line) and phonons (blue dashed line) as a function of the pump parameter, $\Theta$, for $g_{cm}=0.02$ and $\bar{n}_{th}=0$. Vertical lines indicate the interaction times where the number of photons and phonons is restricted to $k=0$ (green dotted line) and for photons to $k=1$ (magenta dotted line). The insets show clearly the corresponding Fock states at these interaction times. $(b)$ Phonon second-order coherence degree as a function of the pump parameter and optomechanical couplings for $\bar{n}_{th}=0.01$. Here we compare: $(i)$ analytical Eq. \ref{g2} (solid) and $(ii)$ numerical solution of the Eq. \ref{me} (dotted). Other parameters (in units of $\omega_m$) are: $r=80$, $g_{ac}=3$, $\delta=0$, $\alpha=0.3$, $\kappa_{a}=3$, $\kappa_{b}=0.05$.}
\label{Fig4}
\end{figure*}
%%%%%%%%%%%%%%%%%%%%%%%%%%%%%%%
Next, we can evaluate a complete analytical solution for the steady state elements of the cavity density matrix. This is because the ME leads to a dynamics, in which the diagonal and off-diagonal elements in the Fock states basis are decoupled. Finally, the detailed balance approach leads to the following expression for the diagonal elements \cite{Filipowicz}
\begin{equation}
P_{n}=N\prod_{k=1}^{\infty}\frac{r}{\kappa_{a}}\frac{\sin^{2}(g_{ac}\tilde{\Theta}\sqrt{k/\omega_{m}r})}{k}
\end{equation}
where $N$ is a constant determined by normalization. The photon trapping condition occurs for certain specific values of $\tilde{\Theta}$ such that, $\sin{\left(g_{ac}\tilde{\Theta}\sqrt{\frac{k+1}{\omega_{m}r}}\right)}=0$ with $k=0,1,2,\ldots$, then $P_{n}=0$ for $n>k$. So, the matrix element that generates transitions between the $k$ and $k+1$ number states vanishes, e.g. for vacuum photon trapping state, $k=0$, one finds that $\tilde{\Theta}=\left\{\tilde{\Theta}_{1}\approx9.36, \tilde{\Theta}_{2}\approx18.73, \tilde{\Theta}_{3}\approx28.09\right\}$, almost same values as for $\{\Theta_1, \Theta_2, \Theta_3\}$, see Fig.\ref{Fig2}.

In Fig. \ref{Fig3}(a) it is shown the behavior of the steady state average number of excitations for both modes, photons and phonons, as functions of the pump parameter, $\Theta$. We can see that the vacuum trapping states for both modes exist at almost same interaction times. Certainly the steady states photon average number reach lower values than the phonon states at the minima, since between the photons and the atoms the interaction is direct and result strongly nonlinear, while the phonons interact with the atoms through the cavity field, so the effect of phonon trapping state is weaker.

As observed in Fig. \ref{Fig3}(b), the optimal synchronization is achieved for a relatively low field amplitude, i.e. $\vert \alpha\vert \ll1$. For more clarity, in Fig. \ref{Fig3}(c) we show that e.g. for the interaction times, $\Theta_{1}$ and $\Theta_{2}$, the Wigner functions for both subsystems look similar, corresponding to the vacuum states. For other interaction times these functions show different states. In the case of the cavity density operator, one gets a superposition of the Fock states while for the MO one has a displaced coherent state. This last result is expected from Eq. \ref{gen} for values where the interaction time leads to $B(\tau)\neq0$.

%%%%%%%%%%%%%%%%%%%%%%%%%%%%%%%%%%%%%%%%%%%%%%%%%%%
\subsection{\label{sec:level6}Second-order correlation function}
%%%%%%%%%%%%%%%%%%%%%%%%%%%%%%%%%%%%%%%%%%%%%%%%%%%
In this section, we evaluate one important characteristics of the micromaser model, that corresponds to the degree of coherence of the maser emission, which can be quantified by the second-order correlation function, $g^{(2)}(0)$. 

In Fig. \ref{Fig4}(a) we present the numerical calculations for the cavity and MO second-order correlation functions vs. the pump parameter, $\Theta$, for the baths at zero temperature. As result, the photons and phonons evidence a super-Poissonian statistics for the interaction times where the trapping of vacuum states occur, see vertical green dotted line. Additionally, we can find that the cavity evidences the photon blockade effect $(g_{a}^{(2)}(0)\rightarrow0)$, matching perfectly to the trapping of one photon, see vertical magenta dotted line and respectively the inset showing the generation of one photon state. Therefore a correspondence between the two phenomena is observed for the one-photon case, however there is no correspondence between the blockade and trapping for the vacuum states of photons and phonons. On the other hand, as will be explained in the next section, one may realize a matching between the blockade and trapping corresponding to the phonon vacuum state.

In the following, we calculate analytically the mechanical second-order correlation function defined by $g_{b}^{(2)}(0)=\langle\hat{b}^{\dagger}\hat{b}^{\dagger}\hat{b}\hat{b}\rangle/\langle\hat{b}^{\dagger}b\rangle^{2}$, and after a simple, but rather long calculation (see details in Appendix \ref{apendice5}) one gets the final expression
\begin{equation}\label{g2}
    g_{b}^{(2)}(0)=\frac{2\bar{n}^{2}_{th}+4\beta^{2}_{1}\bar{n}_{th}+\beta^{4}_{1}}{\bar{n}^{2}_{th}+2\beta^{2}_{1}\bar{n}_{th}+\beta^{4}_{1}}.
\end{equation}
Now, let's study the effect of the optomechanical coupling on the MO connected to a thermal bath at finite temperature. Then, in Fig. \ref{Fig4}(b) one shows the mechanical second-order correlation function as a function of the pump parameter, $\Theta$, for different optomechanical couplings. The environment temperature has been set such that $\bar{n}_{th}=0.01$. The solid lines correspond to the analytical solutions, Eq. \ref{g2}, and the dashed lines correspond to the numerical solutions, Eq. \ref{me}. In particular, one observes when there is no optomechanical interaction, i.e. $g_{cm}=0$, the phonon correlation function evidences a thermal distribution: $g_{b}^{(2)}(0)=2, \forall\Theta$. This result can be deduced from Eq. \ref{rom} since $\hat{\rho}_{m}(\tau)=\hat{\rho}_{m}(0)$ for the null optomechanical coupling and the MO is initially in equilibrium with the thermal bath. On the other hand, the optomechanical coupling will stimulate the coherent phonon statistics, i.e. $g_{b}^{(2)}(0)\sim1$ outside of the specific values $\left\{\Theta_{1},\Theta_{2},\Theta_{3}\right\}$ corresponding to phonon trapping states. However, in the close vicinity of these values the phonon statistics is super-Poissonian. 

As conclusion of this section, there is no sign of a blockade effect in the case of a thermal phonon bath for any pump parameter, only a coherent phonon emission stimulated by the optomechanical coupling is observed.
%%%%%%%%%%%Figure 5%%%%%%%%%%%%%%
\begin{figure*}[t]
\centering
\includegraphics[width=0.247\linewidth]{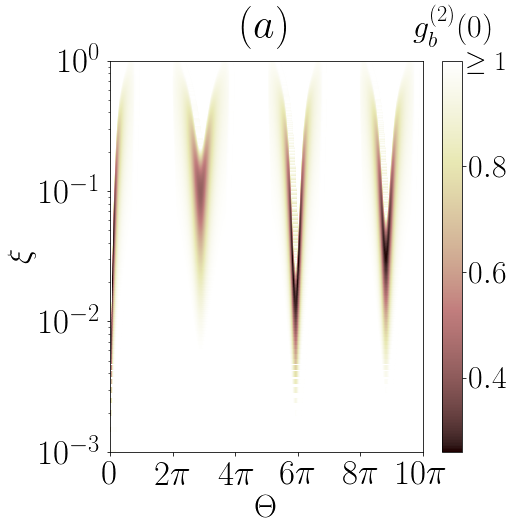}
\hspace{0.05cm}
\includegraphics[width=0.237\linewidth]{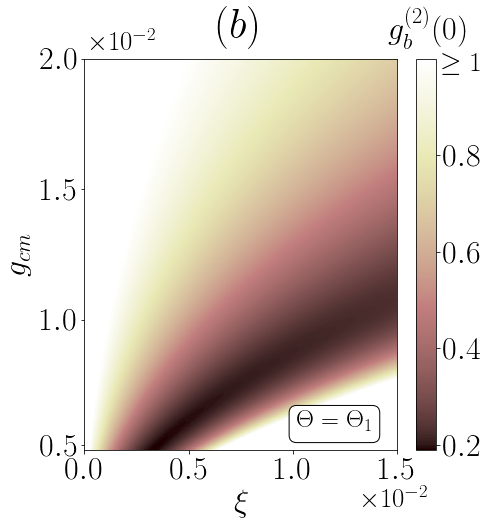}
\hspace{0.05cm}
\includegraphics[width=0.237\linewidth]{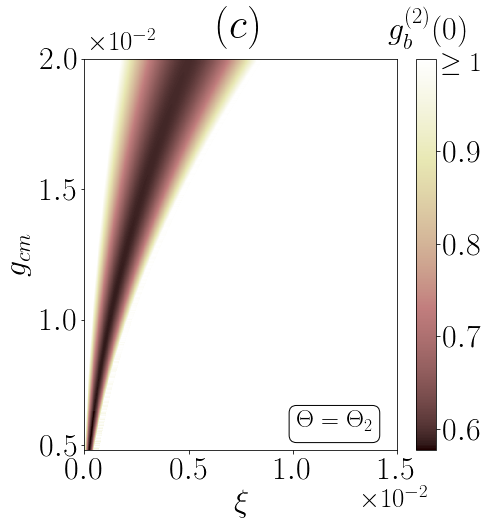}
\hspace{0.05cm}
\includegraphics[width=0.237\linewidth]{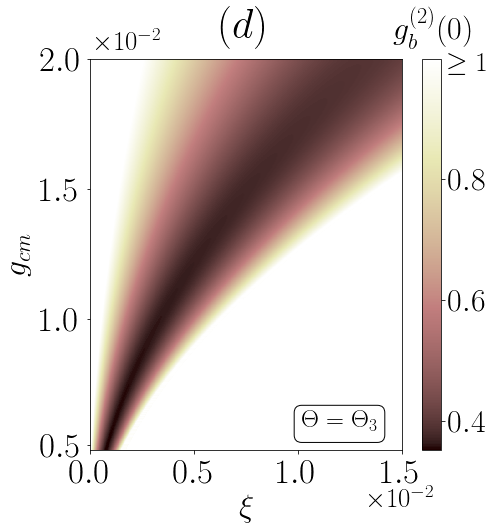}
\caption{$(a)$ Second-order coherence degree as a function of the squeezing parameter, $\xi$ and the pump parameter, $\Theta$ (with $g_{cm}=0.014$). Second order coherence degree as a function of the optomechanical coupling, $g_{cm}$ and the squeezing parameter, $\xi$, for the pump parameter: $(b)$ $\Theta=\Theta_{1}$, $(c)$ $\Theta=\Theta_{2}$, $(d)$ $\Theta=\Theta_{3}$, obtained from the phonon trapping condition Eq. \ref{cond}. As it is observed, there is an optomechanical coupling that optimizes the phonon blockade effect for a given $\xi$, according to the defined pump parameter.  Other parameters (in units of $\omega_m$) are: $g_{ac}=3$, $\delta=0$, $\alpha=0.3$, $r=80$, $\kappa_{b}=0.05$, $\phi=\pi$ and $\bar{n}_{th}=0$.} \label{Fig5}
\end{figure*}
%%%%%%%%%%%%%%%%%%%%%%%%%%%%%%%

%%%%%%%%%%%%%%%%%%%%%%%%%%%%%%%%%%%%%%%%%%%%%%%%%%%
\section{\label{sec:level7}Phonon blockade with squeezed vacuum reservoir}
%%%%%%%%%%%%%%%%%%%%%%%%%%%%%%%%%%%%%%%%%%%%%%%%%%%
In the following we discuss how to realize the effect of phonon blockade and its relation to the trapping states. As was mentioned in the Introduction, to get the phonon blockade, according to the recent studies \cite{Yang2020, Lin2021, Restrepo2017, Liu2010, Xu2019, APL2021} two mechanisms can be used: i) conventional - by a strong driven phonon nonlinearity in the system, or ii) unconventional - by the destructive interference between different paths which restrict the higher than one-phonon excitations.

In the present model we propose the enhancement of the phonon nonlinearity by considering the interaction of the system with a phonon vacuum squeezed reservoir. Therefore, the MO dynamics is described by the following maser ME
\begin{eqnarray}\label{dinamica}
    \frac{d\hat{\rho}_{m}}{dt}
    &=&
    r\left(\hat{\mathcal{M}}_{m}(\tau)-1\right)\hat{\rho}_{m}\nonumber\\
    &+&
    \frac{\kappa_{b}}{2}(N_{sq}+1)\left(2\hat{b}\hat{\rho}_{m} \hat{b}^{\dagger}- \hat{b}^{\dagger}\hat{b}\hat{\rho}_{m}-\hat{\rho}_{m} \hat{b}^{\dagger}\hat{b}\right)\nonumber\\
    &+&
    \frac{\kappa_{b}}{2}N_{sq}\left(2\hat{b}^{\dagger}\hat{\rho}_{m} \hat{b}-\hat{b}\hat{b}^{\dagger}\hat{\rho}_{m}-\hat{\rho}_{m} \hat{b}\hat{b}^{\dagger}\right)\nonumber\\
    &+&
    \frac{\kappa_{b}}{2}M_{sq}\left(2\hat{b}^{\dagger}\hat{\rho}_{m} \hat{b}^{\dagger}- \hat{b}^{\dagger}\hat{b}^{\dagger}\hat{\rho}_{m}-\hat{\rho}_{m} \hat{b}^{\dagger}\hat{b}^{\dagger}\right)\nonumber\\
    &+&
    \frac{\kappa_{b}}{2}M_{sq}^{*}\left(2\hat{b}\hat{\rho} \hat{b}-\hat{b}\hat{b}\hat{\rho}-\hat{\rho} \hat{b}\hat{b}\right),
\end{eqnarray}
where $N_{sq}=\sinh^{2}{\xi}$ corresponds to the average number of phonons in the squeezed reservoir at zero temperature and the quantity $M_{sq}=-\exp{(i\phi)}\sinh{\xi}\cosh{\xi}$, obeys the relation $\sqrt{N_{sq}(N_{sq}+1)}=\vert M_{sq}\vert $. Here, the parameters $\xi$ and $\phi$ represent the squeezing amplitude and the phase, respectively, as appear in the definition of the complex squeezing parameter, $\zeta=\xi\exp{[i\phi]}$.

In Fig. \ref{Fig5}(a) we plot the phonon second-order coherence degree as a function of the squeezing parameter, $\xi$, and pump parameter, $\Theta$. On the one hand, for the no squeezing case, $\xi=0$, super-Poissonian peaks occur, i.e $g_{b}^{(2)}(0)>1$. On the other hand, for the squeezed reservoir with $\xi\sim [0.01,1]$, we find that the phonon correlation function evidences sub-Poissonian phonon statistics, i.e. $g_{b}^{(2)}(0)<1$, for specific values of the pump parameter. In addition, one observes that if coming closer to the vacuum trapping state, the stronger blockade effect $(g_{b}^{(2)}(0)\rightarrow0)$ can be achieved with less squeezing, $\xi\sim10^{-2}$.

In Fig. \ref{Fig5}(b-d) we plot the phonon second-order coherence degree as a function of the optomechanical coupling, $g_{cm}$ and the squeezing parameter, $\xi$, for each value of pump parameter, $\Theta$, where the phonon vacuum trapping states occur. As result, one finds that in each of these $\Theta$ values, there is an optomechanical coupling that optimizes the phonon blockade according to a certain squeezing parameter. 

In Fig. \ref{Fig6} we plot the phonon second-order correlation function (left vertical axis) and the steady state average number of phonons and photons (right vertical axis) vs. pump parameter, $\Theta$. One finds that the phonon correlation function evidences a phonon blockade ($g_{b}^{(2)}(0)<1$) at the positions where phonon and photon vacuum trapping states are generated. In the remaining intervals of $\Theta$, it is a coherent distribution, $g_{b}^{(2)}(0)=1$, i.e. a phonon lasing effect. Since we are in a good synchronization zone, $\alpha=0.384$ (see Fig. \ref{Fig3}b), both trapping states of phonons and photons, represent a good witness for the phonon blockade effect.

Therefore, the generation of the phonon blockade effect in HMM and its detection by the photon and phonon trapping vacuum states conclude this section and these effects represent the main results of the work.

%%%%%%%%%%%%%%%%%%%%%%%%%%%%%%%%%%%%%%%%%%%%%%%%%%%
\section{\label{sec:level8}Discussion}
%%%%%%%%%%%%%%%%%%%%%%%%%%%%%%%%%%%%%%%%%%%%%%%%%%%
In summary, we have proposed a hybrid micromaser model consisting of atoms passing at a given rate through a cavity with a moving mirror, which is connected to a low temperature thermal bath in one case or to a vacuum squeezed phonon reservoir in another case. We demonstrate, both analytically and numerically, that for certain interaction times between the atoms and the optomechanical cavity, vacuum phonon trapping states are generated. 
In this framework, with independent phonon and photon thermal baths, we show that the trapping of the phonons and photons may occur simultaneously, i.e. at same value of $\Theta$ as observed in Figs. \ref{Fig3}(a,c). Therefore, we find that for a low coherence amplitude of the initial cavity state one has better synchronization of trapping of the phonon and photon states, see Fig. \ref{Fig3}(b). Moreover, by increasing the optomechanical coupling, the phonon second-order correlation function evidences the effect of a coherent distribution, $g_{b}^{(2)}(0)\sim1$, that occurs for the finite steady state phonon number, i.e. outside of the vacuum trapping states, see Fig. \ref{Fig4}(b). 

\begin{figure}[t]
\centering
\includegraphics[width=0.9\linewidth]{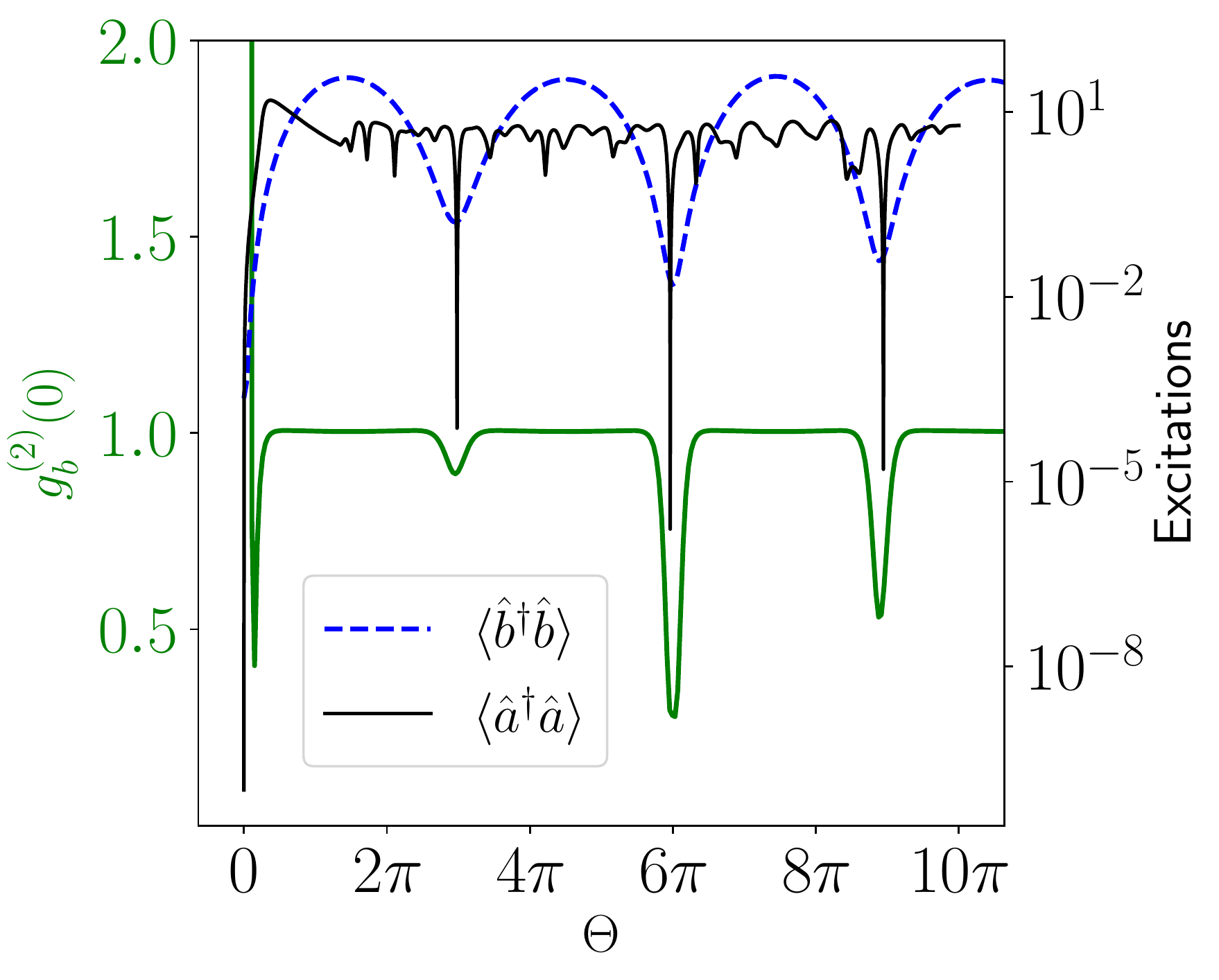}
\caption{Second-order correlation function for phonons (green curve) and steady state average number of photons (black curve) and phonons (blue curve) vs. the pump parameter, $\Theta$. Other parameters (in units of $\omega_m$) are: $g_{ac}=3$, $g_{cm}=0.014$, $\delta=0$, $\alpha=0.384$, $\kappa_{a}=3$, $\kappa_{b}=0.05$, $r=80$, $\xi=0.015$, $\phi=\pi$, $\bar{n}_{th}=0$.}
\label{Fig6}
\end{figure}

Besides that the creation of the phonon trapping states \textit{per se} is an interesting and new (to the best of our knowledge) quantum effect in a hybrid maser architecture, our work goes further and we propose this effect as a witness and control of another important quantum effect that is the phonon blockade (see Fig. \ref{Fig6}). To have a photon/phonon blockade effect, commonly one needs a strong photon/phonon nonlinearity in the system. However, as our model of HMM is developed in the weak optomechanical coupling regime with the mechanical mode connected to a thermal bath, all this will result in a very weak phonon nonlinearity and hence the phonon blockade is not observed, see Fig. \ref{Fig4}(a-b). At the same time the photon nonlinearity is strong via the Jaynes-Cummings coupling, and so the photon blockade may occur for some values of the atom-cavity interaction time, see in Fig. \ref{Fig4}(a) when $g_a^{(2)}(0)<1$. 
Therefore, we propose to connect the HMM to a vacuum squeezed phonon reservoir in order to search the phonon blockade. As result, one observes that in the close neighborhood of the critical points of $\Theta$, resulted from Eq. \ref{cond}, where the trapping states occur, the phonon distribution becomes sub-Poissonian so that $g_{b}^{(2)}(0)<1$, corresponding to so-called phonon blockade effect (see Figs. \ref{Fig5} - \ref{Fig6}). 

From the experimental point of view the model of HMM can in principle use a similar setup of the standard MM, where the photon trapping effect was observed for the first time \cite{Weidinger}. To produce the optomechanical interaction, the microwave cavity will have a movable mirror. The configuration of a FPC with oscillatory mirror could be inspired from the detectors of the gravitational waves (GW), where the optomechanical coupling is the main ingredient and very well studied \cite{Aspelmeyer}. Of course, the GW interferometers usually are of very large dimensions as compared to the common FPC used for maser and laser setups, but what one needs is particularly to consider the physical functionality of the cavity with a moving mirror. There exist some setups of small-sized tabletop GW interferometers using Fabry Perot resonator \cite{GWD}, which may serve as desirable examples to develop an experimental prototype of HMM optimal to observe the phonon trapping and blockade effects. Moreover our proposal is developed in the regime of weak optomechanical coupling, hence the HMM does not require a strong coupling between the cavity and vibrating mirror. Additionally to point out, in the standard MM experiments one finds a particular problem, if the cavity is crossed simultaneously by two atoms, the photon trapping leaks and eventually get destroyed \cite{Orszag94}. This effect causes an experimental difficulty, since the atoms arrive to the cavity at random times and there is a non zero probability of having such a situation. In the case of phonon HMM, we may encounter a similar difficulty, unless the atoms pass through a very fine velocity filter.

As a central result of the present study, we identify an interconnection between the trapping and blockade effects for phonons and photons in the HMM. Particularly if one approaches to the phonon trapping states, which additionally is synchronized to the photon trapping, the phonon blockade effect can be achieved for a slightly squeezed reservoir and corresponding optomechanical coupling, see Fig. \ref{Fig5}. Our proposal for witnessing the phonon blockade by the trapping states in a hybrid micromaser setup could facilitate the experimental detection of the phonon blockade effects. In conclusion, we can realize in the HMM setup several common maser properties for the phonon and photon modes, like coherence, trapping and sub-Poissonian states. These effects highlight the findings of this work.

%%%%%%%%%%%Figure 6%%%%%%%%%%%%%%
%%%%%%%%%%%%%%%%%%%%%%%%%%%%%%%

\begin{acknowledgments}
H.M. acknowledge financial support from Universidad Mayor through the Doctoral fellowship. V.E. and M.O. acknowledge the financial support from Fondecyt Regular No. $1180175$. V.E. acknowledge grant No. $20.80009.5007.01$ of the State Program (2020-2023) from National Agency for Research and Development of Moldova. 
\end{acknowledgments}

\appendix

\section{\label{apendice1}Hamiltonian in the interaction picture}
%%%%%%%%%%%%%%%%%%%%%%%%%%%%%%%%%%%%%%%%%%%%%%%%%%%
This appendix provides the derivation of the initial Hamiltonian Eq. \ref{base} in the interaction picture. Let us introduce the operator counting the number of atom-cavity polaritonic excitations:

\begin{equation}
    \hat{\mathcal{N}}=\hat{a}^{\dagger}\hat{a}+\hat{\sigma}_{z}/2
\end{equation}
We observe that this polariton number operator commutes with
the Hamiltonian of the system, $[\hat{\mathcal{H}},\hat{\mathcal{N}}]=0$. Therefore,
the Hamiltonian of the closed system is block-diagonal in the
basis of eigenvectors of the polariton number operator. By considering the detuning $\delta=\omega_{a}-\omega_{c}$, the Hamiltonian in Eq. \ref{base} can be written as
\begin{equation}
   \hat{\mathcal{H}}=\hat{\mathcal{H}}_{0}+\hat{\mathcal{H}}_{I}, 
\end{equation}
where
\begin{align}
    \hat{\mathcal{H}}_{0}&=\omega_{m}\hat{b}^{\dagger}\hat{b},\nonumber\\ 
   \hat{\mathcal{H}}_{I}&=\delta\frac{\hat{\sigma}_{z}}{2}+
   g_{ac}\left(\hat{a}\hat{\sigma}_{+}+\hat{a}^{\dagger}\hat{\sigma}_{-}\right)
   -g_{cm}\hat{a}^{\dagger}\hat{a}\left(\hat{b}^{\dagger}+\hat{b}\right).
\end{align}
Now, we calculate the Hamiltonian in the first interaction picture, that is
\begin{equation}\label{cuadro1}
    \hat{\mathcal{H}}'=e^{i \hat{\mathcal{H}}_{0} t}\hat{\mathcal{H}}_{I}e^{-i \hat{\mathcal{H}}_{0} t}.
\end{equation}
Using the fact that
\begin{equation}
    \mathcal{\hat{U}}f\left(\left\{\mathcal{\hat{X}}_{i}\right\}\right)\mathcal{\hat{U}}^{\dagger}
    =f\left(\left\{\mathcal{\hat{U}}\mathcal{\hat{X}}_{i}\hat{U}^{\dagger}\right\}\right),
\end{equation}
for any function $f$, unitary operator $\mathcal{\hat{U}}$ and arbitrary set of operators $\left\{\mathcal{\hat{X}}_{i}\right\}$, so the Eq. \ref{cuadro1} takes the form
\begin{eqnarray}
    \hat{\mathcal{H}}'=\hat{\mathcal{H}}'_{0}+\hat{\mathcal{H}}'_{I},
\end{eqnarray}
where
\begin{align}
    \hat{\mathcal{H}}'_{0}&=\delta\frac{\hat{\sigma}_{z}}{2}+
   g_{ac}\left(\hat{a}\hat{\sigma}_{+}+\hat{a}^{\dagger}\hat{\sigma}_{-}\right),\nonumber\\ 
   \hat{\mathcal{H}}'_{I}&=-g_{cm}\hat{a}^{\dagger}\hat{a}\left(\hat{b}^{\dagger}e^{i\omega_{m}t}+\hat{b}e^{-i\omega_{m}t}\right).
\end{align}
Now, we move to a second interaction picture, defined as
\begin{equation}
    \hat{\tilde{\mathcal{H}}}=\exp{\left[i\int\hat{\mathcal{H}}'_{0}dt\right]}\hat{\mathcal{H}}'_{I}\exp{\left[-i\int\hat{\mathcal{H}}'_{0}dt\right]}.
\end{equation}
The result of the transformation is easily calculated to be finally the Eq. \ref{model}.\\

%%%%%%%%%%%%%%%%%%%%%%%%%%%%%%%%%%%%%%%%%%%%%%%%%%%
\section{\label{apendice2}Phonon gain superoperator}
%%%%%%%%%%%%%%%%%%%%%%%%%%%%%%%%%%%%%%%%%%%%%%%%%%%
This section presents the derivation of the gain part of the maser ME for the MO operator. We consider an initial atom-cavity-oscillator operator as
\begin{equation}
\hat{\rho}(0)=\begin{pmatrix}
0 & 0\\\\
0 & 1
\end{pmatrix}\otimes\hat{\rho}_{c}(0)\otimes\hat{\rho}_{m}(0),
\end{equation}
where the atom was taken in the ground state. After the interaction time $\tau$, the total density operator evolves according to Eq. \ref{unitary}. After a straightforward calculation, one gets
\begin{widetext}
\begin{equation}
    \hat{\rho}(\tau)=
\begin{pmatrix}
g_{ac}^{2}\mathcal{\mathcal{\hat{S}}}\hat{a}e^{i\mathcal{\hat{F}}}\hat{\rho}_{c}(0)\otimes\hat{\rho}_{m}(0)e^{-i\mathcal{\hat{F}}}\hat{a}^{\dagger}\mathcal{\mathcal{\hat{S}}} & -i g_{ac}\mathcal{\mathcal{\hat{S}}}\hat{a}e^{i\mathcal{\hat{F}}}\hat{\rho}_{c}(0)\otimes\hat{\rho}_{m}(0)\mathcal{\hat{D}}^{\dagger}\\\\
 ig_{ac}\mathcal{\hat{D}}\hat{\rho}_{c}(0)\otimes\hat{\rho}_{m}(0)e^{-i\mathcal{\hat{F}}}\hat{a}^{\dagger}\mathcal{\mathcal{\hat{S}}} & \mathcal{\hat{D}}\hat{\rho}_{c}(0)\otimes\hat{\rho}_{m}(0)\mathcal{\hat{D}}^{\dagger} 
\end{pmatrix},
\end{equation}
\end{widetext}
where $\mathcal{\hat{D}}^{\dagger}$ is hermitian conjugate of $\mathcal{\hat{D}}$, see Eq. \ref{eq5}. 

Next, tracing over the atom, the cavity-MO reduced operator reads
\begin{equation}
    \hat{\rho}_{c,m}(\tau)
    =
    \hat{\rho}_{m}(0)\mathcal{\hat{D}}\hat{\rho}_{c}(0)\mathcal{\hat{D}}^{\dagger}+
    g^{2}_{ac}e^{i\mathcal{\hat{F}}}\hat{\rho}_{m}(0)e^{-i\mathcal{\hat{F}}}\mathcal{\mathcal{\hat{S}}}\hat{a}\hat{\rho}_{c}(0)\hat{a}^{\dagger}\mathcal{\mathcal{\hat{S}}}.
\end{equation}
In the following, we consider the cavity field initialized in a coherent state (Eq. \ref{coh}). Hence, the above equation becomes
\begin{eqnarray}
    \hat{\rho}_{c,m}(\tau)
    &=&
    \hat{\rho}_{m}(0)e^{-\vert \alpha\vert ^{2}}\sum_{n,m}\frac{\alpha^{n}\alpha^{*m}}{\sqrt{n!m!}}D_{n}D_{m}^{*}\vert n\rangle\langle m\vert \nonumber\\
    &+&
    g^{2}_{ac}e^{i\mathcal{\hat{F}}}\hat{\rho}_{m}(0)e^{-i\mathcal{\hat{F}}}e^{-\vert \alpha\vert ^{2}}\sum_{n,m}\frac{\alpha^{n}\alpha^{*m}}{\sqrt{n!m!}}\sqrt{nm}\nonumber\\
    &\times&
    S_{n-1}S_{m-1}\vert n-1\rangle\langle m-1\vert ,
\end{eqnarray}
where
\begin{align}
    D_{n}&=\cos{\left(\sqrt{\varphi_{n}}\tau\right)}+\frac{i\delta}{2}\frac{\sin{\left(\tau\sqrt{\varphi_{n}}\right)}}{\sqrt{\varphi_{n}}},\\
    S_{n-1}&=\frac{\sin{\left(\tau\sqrt{\varphi_{n-1}+g_{ac}^{2}}\right)}}{\sqrt{\varphi_{n-1}+g_{ac}^{2}}}.
\end{align}
Finally tracing over the cavity, we find the Eqs. \ref{rom}, \ref{coefab} and \ref{coefab2}.

%%%%%%%%%%%%%%%%%%%%%%%%%%%%%%%%%%%%%%%%%%%%%%%%%%%
\section{\label{apendice3}Solution of the Fokker-Planck equation}
%%%%%%%%%%%%%%%%%%%%%%%%%%%%%%%%%%%%%%%%%%%%%%%%%%%
Here we present the steps to get the solution \ref{sfp} of the Fokker-Planck equation \ref{fp}. Let us consider the MO initially is in a coherent state, i.e. in the Gaussian representation takes the form
\begin{equation}
    P(\beta,\beta^{*},0)=(\pi\epsilon)^{-1}\exp{\left[-\vert \beta-\beta_{0}\vert ^{2}/\epsilon\right]}
\end{equation}
By considering a solution type $P(\beta,\beta^{*},t)=\exp{\left[a(t)+b(t)\beta+c(t)\beta^{*}+d(t)\beta\beta^{*}\right]}$ one obtains a set of first order differential equations
\begin{align}
    \dot{a}(t)&=\kappa_{b}\left(d(t)+\bar{n}_{th}d^{2}(t)\right),\\
    \dot{b}(t)&=\kappa_{b}\left(\frac{1}{2}c(t)+\bar{n}_{th}c(t)d(t)\right)-\lambda B(\tau)rd(t),\\
    \dot{c}(t)&=\kappa_{b}\left(\frac{1}{2}b(t)+\bar{n}_{th}b(t)d(t)\right)-\lambda B(\tau)rd(t),\\
    \dot{d}(t)&=\kappa_{b}\left(1+\bar{n}_{th}[c(t)b(t)+d(t)]\right)-\lambda B(\tau)r[c(t)+b(t)].
\end{align}
For an initial thermal distribution, $\epsilon=\bar{n}_{th}$, $\beta_{0}=0$ and after a straightforward calculation, we get 

\begin{eqnarray}
    P(\beta,\beta^{*},t)
    &=&
    \frac{1}{\pi\bar{n}_{th}}\textrm{exp}[\textrm{ln}[\pi\bar{n}_{th}]\nonumber\\
    &-&\frac{\beta_{1}^{2}}{\bar{n}_{th}}-\frac{|\beta|^{2}}{\bar{n}_{th}}+\frac{\beta_{1}\beta}{\bar{n}_{th}}+\frac{\beta_{1}\beta^{*}}{\bar{n}_{th}}],
\end{eqnarray}

with $\beta_{1}=2\lambda r  B(\tau)\kappa_{b}^{-1}(1-\exp{\left[-\kappa_{b}t/2\right]})$. Finally, the last equation can be written compactly as Eq. \ref{sfp}.
%%%%%%%%%%%%%%%%%%%%%%%%%%%%%%%%%%%%%%%%%%%%%%%%%%%
\section{\label{apendice4}Phonon trapping condition}
%%%%%%%%%%%%%%%%%%%%%%%%%%%%%%%%%%%%%%%%%%%%%%%%%%%
The critical values of the pump parameter, $\Theta$, where the phonon trapping vacuum states occur, are calculated from the condition of the minimum value for the steady state average phonon number: $\frac{\partial}{\partial\Theta}\langle \hat{b}^{\dagger}\hat{b}\rangle=0$ with $\frac{\partial^{2}}{\partial\Theta^{2}}\langle \hat{b}^{\dagger}\hat{b}\rangle>0$. So the derivative of Eq. \ref{analitico} gives:
%\begin{widetext}
\begin{eqnarray}
    \frac{\partial}{\partial\Theta}\langle \hat{b}^{\dagger}\hat{b}\rangle
    &=&
    \frac{8e^{-2\vert \alpha\vert ^{2}}g_{ac}\lambda^{2}r^{3/2}}{\kappa_{b}^{2}\omega_{m}^{1/2}}\nonumber\\
    &\times&
    \sum_{n}\frac{\vert \alpha\vert ^{2(n+1)}\sqrt{n+1}}{(n+1)!}\sin{\left(2g_{ac}\Theta\sqrt{\frac{n+1}{\omega_{m}r}}\right)}\nonumber\\
    &\times&
    \sum_{n}\frac{\vert \alpha\vert ^{2(n+1)}}{(n+1)!}\sin{\left(g_{ac}\Theta\sqrt{\frac{n+1}{\omega_{m}r}}\right)}^{2}.
\end{eqnarray}
%\end{widetext}
Taking into account that $\left\{\alpha,g_{ac},r,\lambda,\kappa_{b}\right\}>0$, the phonon trapping state is valid for
\begin{equation}
    \sum_{n}\frac{\vert \alpha\vert ^{2(n+1)}\sqrt{n+1}}{(n+1)!}\sin{\left(2g_{ac}\Theta\sqrt{\frac{n+1}{\omega_{m}r}}\right)}=0.
\end{equation}
Finally, the above equation can be solved numerically, and the minimal values for $\langle\hat{b}^{\dagger}\hat{b}\rangle$ are found, see vertical lines in Fig. \ref{Fig2}$(b)$.

%%%%%%%%%%%%%%%%%%%%%%%%%%%%%%%%%%%%%%%%%%%%%%%%%%%
\section{\label{apendice5}Calculation of $g_{b}^{(2)}(0)$ % using the moment-generating function $\mathcal{Q}(s)$
}
%%%%%%%%%%%%%%%%%%%%%%%%%%%%%%%%%%%%%%%%%%%%%%%%%%%
To calculate the second-order correlation function for phonons, $g_{b}^{(2)}(0)=\langle\hat{b}^{\dagger}\hat{b}^{\dagger}\hat{b}\hat{b}\rangle/\langle\hat{b}^{\dagger}b\rangle^{2}$, we consider convenient to use the moment-generating function
\begin{equation}
    \mathcal{Q}(s)=\sum_{n=0}^{\infty}(1-s)^{n}P(n),
\end{equation}
where $P(n)=\int d^{2}\beta P(\beta,\beta^{*},t)\vert \langle n\vert \beta\rangle\vert ^{2}$ is the probability to have $n$ phonons in the MO and $P(\beta,\beta^{*},t)$ is defined in Eq. \ref{sfp}. After a straightforward calculation one obtains
\begin{equation}
    P(n)=
    \frac{1}{\pi\bar{n}_{th}n!} \int d^{2}\beta \vert \beta\vert ^{2n} e^{-\vert \beta\vert ^{2}-\frac{\vert \beta-\beta_{1}\vert ^{2}}{\bar{n}_{th}}}.
\end{equation}
Finally, using the definition $g^{(2)}(0)=\frac{1}{\langle n\rangle}\frac{d^{2}\mathcal{Q}}{ds^{2}}\vert _{s=0}$, see \cite{Scully}, we get the Eq. \ref{g2}.

\bibliography{apssamp}

\end{document}